\documentclass[11pt]{article}
\usepackage{amsmath,amssymb,amsthm} 
\usepackage{bbm}
\usepackage{stmaryrd}
\usepackage{paralist}
\usepackage{hyperref}
\usepackage{color}

\setdefaultenum{(i)}{}{}{}
\usepackage[margin=0.99in]{geometry} 
\usepackage[pdftex]{graphicx}
\usepackage{subfigure}
\theoremstyle{plain}
\newtheorem{mythm}{Theorem} \numberwithin{mythm}{section}

\newtheorem{myrek}[mythm]{Remark}

\DeclareMathAlphabet\scr{U}{scr}{m}{n}
\SetMathAlphabet\scr{bold}{U}{scr}{b}{n}
  \DeclareFontFamily{U}{scr}{\skewchar\font'177}%
  \DeclareFontShape{U}{scr}{m}{n}{<-6>rsfs5<6-8>rsfs7<8->rsfs10}{}%
  \DeclareFontShape{U}{scr}{b}{n}{<-6>rsfs5<6-8>rsfs7<8->rsfs10}{}%

\numberwithin{equation}{section}

\renewenvironment{thebibliography}[1]{%
\begin{oldthebibliography}{#1}%
\setlength{\baselineskip}{.8em}
\linespread{.8}
\small
\setlength{\parskip}{0ex}%
\setlength{\itemsep}{.1em}%
}%
{%
\end{oldthebibliography}%
}

\DeclareMathAlphabet\scr{U}{scr}{m}{n}
\SetMathAlphabet\scr{bold}{U}{scr}{b}{n}
  \DeclareFontFamily{U}{scr}{\skewchar\font'177}%
  \DeclareFontShape{U}{scr}{m}{n}{<-6>rsfs5<6-8>rsfs7<8->rsfs10}{}%
  \DeclareFontShape{U}{scr}{b}{n}{<-6>rsfs5<6-8>rsfs7<8->rsfs10}{}%

\newcommand{\E}{\scr E}

\newcommand{\mal}{\stackrel{\mbox{\tiny$\bullet$}}{}}

\numberwithin{equation}{section}

\begin{document}

\title{\vspace{-1.5cm}The General Structure of Optimal Investment\\ and Consumption with Small Transaction Costs\footnote{The authors are grateful to Ale{\v{s}} {\v{C}}ern{\'y}, Christoph Czichowsky, Paolo Guasoni, Ren Liu, Richard Martin, Marcel Nutz, Walter Schachermayer, Torsten Sch\"oneborn, Steven E.\ Shreve, Mihai S\^irbu, H.\ Mete Soner, and Nicholas Westray for fruitful discussions. They also thank two anonymous referees and the editor for numerous pertinent remarks.}}
\author{
Jan Kallsen
\thanks{Christian-Albrechts-Universit\"at zu Kiel, Mathematisches Seminar, Westring 383, D-24098 Kiel, Germany, email \texttt{kallsen@math.uni-kiel.de}. Financial support by DFG research grant KA 1682/4-1 is gratefully acknowledged.}
\and
Johannes Muhle-Karbe\thanks{ETH Z\"urich, Departement f\"ur Mathematik, R\"amistrasse 101, CH-8092, Z\"urich, Switzerland, and Swiss Finance Institute, email \texttt{johannes.muhle-karbe@math.ethz.ch}. Partially supported by the National Centre of Competence in Research ``Financial Valuation and Risk Management'' (NCCR FINRISK), Project D1 (Mathematical Methods in Financial Risk Management), of the Swiss National Science Foundation (SNF), and the ETH Foundation.}
}
\date{}
\maketitle

\maketitle

\begin{abstract}
We investigate the general structure of optimal investment and consumption with small proportional transaction costs. For a safe asset and a risky asset with general continuous dynamics, traded with random and time-varying but small transaction costs, we derive simple formal asymptotics for the optimal policy and welfare.  These reveal the roles of the investors' preferences as well as the market and cost dynamics, and also lead to a fully dynamic model for the implied trading volume. In frictionless models that can be solved in closed form, explicit formulas for the leading-order corrections due to small transaction costs are obtained.
\end{abstract}




\section{Introduction}

Classical financial theory is built on the assumption of perfectly liquid markets. If prices follow continuous-time diffusions, then this also holds for most optimal trading strategies \cite{merton.69,merton.71}. As a result, frictionless models typically prescribe incessant trading, which is unfeasible with even the slightest market imperfections. Proportional transaction costs represent one such friction present even in the most liquid financial markets in the form of bid-ask spreads. The study of their impact on portfolio choice was initiated by the seminal papers of  Constantinides and Magill \cite{magill.constantinides.76, constantinides.86}, as well as Dumas and Luciano \cite{dumas.luciano.91}.\footnote{Cf.\ Taksar et al.\ \cite{taksar.al.88}, Davis and Norman \cite{davis.norman.90}, and Shreve and Soner \cite{shreve.soner.94} for corresponding verification theorems.} In infinite-horizon models with constant risk aversion, transaction costs, and investment opportunities,\footnote{That is, constant expected excess returns and volatilities.} they argued that it is optimal to refrain from trading until one's position leaves a so-called ``no-trade region'' of constant width around the frictionless target. The corresponding welfare effect of transaction costs was found to be small, as ``the derived utility is insensitive to deviations from the optimal portfolio proportions, and investors accommodate large transaction costs by drastically reducing the frequency and volume of trade'' \cite{constantinides.86}. Put differently, investors whose only motive to trade is rebalancing towards a constant frictionless target after substantial price drops or rises do not suffer severely from a wider bid-ask spread if they adjust their trading strategies accordingly.

In the last decades, the substantial decline of bid-ask spreads across financial markets has sparked a huge increase of trading volume, and spurred the continued growth of high-frequency trading. The corresponding high-volume strategies naturally require a careful assessment of the trading costs they generate when reacting dynamically to various signals from the market. Accordingly, there has been growing interest in portfolio choice with transaction costs and stochastic opportunity sets, in financial economics \cite{balduzzi.lynch.99,lynch.balduzzi.00,lynch.tan.11,liu.loewenstein.07,garleanu.pedersen.12,dufresne.al.12}, mathematical finance \cite{soner.touzi.11,kallsen.muhlekarbe.12}, and also in the industry practice of quantitative finance \cite{martin.schoeneborn.11,martin.12,bouchaud.al.12}. Broadly speaking, these studies either employ numerical methods, or study the practically relevant limiting regime of \emph{small} transaction costs to shed more light on the salient features of the solution.

The present study extends and unifies the results of the second strand of research, by abstracting from concrete models and unveiling the general structure of portfolio choice with small transaction costs. Using formal pertuarbation arguments, we derive simple asymptotic formulas for approximately optimal trading strategies as well as the corresponding welfare and implied trading volume in very general settings.\footnote{Corresponding results for the Black-Scholes model have been obtained by Shreve and Soner \cite{shreve.soner.94}, Whalley and Wilmott \cite{whalley.wilmott.97}, Jane{\v{c}}ek and Shreve \cite{janecek.shreve.04}, and in many more recent studies. Using formal perturbation arguments, Martin and Sch\"oneborn \cite{martin.schoeneborn.11,martin.12} study local utility maximizers, and the companion paper of the present study \cite{kallsen.muhlekarbe.12} deals with exponential investors in a general setting. Soner and Touzi \cite{soner.touzi.11} as well as Possama\"i, Soner and Touzi \cite{possamai.al.12} study infinite-horizon consumption models with general utilities in a multidimensional complete market, and provide verification theorems based on the theories of viscosity solutions and homogenization.} We consider investors with general preferences over intermediate consumption and terminal wealth, who receive a random endowment stream and trade a safe and a risky asset with general It\^o process dynamics, in the presence of random and time-varying but small transaction costs. Even in this generality, the structure of the solution has an unexpectedly simple form.

With small costs, investors should keep their holdings in a time- and state dependent no-trade region. The latter is generally not centered around the frictionless target, because past transaction costs reduce investors' wealth. However, the optimal adjustment is the simplest one conceivable: investors just change their target position (and also their optimal consumption) exactly as they would in the frictionless case to account for their reduced wealth. The halfwidth $\Delta \mathrm{NT}_t$ of the optimal no-trade region is given by the cubic root of three factors, stemming from the width of the spread, the optimal frictionless strategy, and the investors' preferences, respectively:
$$
\Delta\mathrm{NT}_t = \left(\frac{3R_t}{2} \frac{d\langle \varphi \rangle_t}{d\langle S \rangle_t} \varepsilon_t\right)^{1/3}.
$$
 Small spreads only enter through their current halfwidth $\varepsilon_t$, i.e., the dynamics of future costs are not hedged at the leading order. For the frictionless optimal strategy $\varphi_t$, the crucial quantity turns is its local quadratic variation $d\langle \varphi \rangle_t$ normalized by the one of the market $d\langle S \rangle_t$, i.e., the ratio of squared diffusion coefficients. The basic tradeoff is that more active target strategies require wider buffers to save transaction costs, whereas turbulent market times call for closer tracking to limit losses due to displacement from the target portfolio. The final ingredient for the width of the no-trade region is the risk tolerance $R_t$ of the investors' indirect utility, which subsumes their preferences by weighting the relative importances of current and future consumption streams against each other.\footnote{Without intermediate consumption, the special case of a risk-tolerance \emph{wealth} process also plays a key role in the work of Kramkov and S{\^{\i}}rbu \cite{kramkov.sirbu.06a,kramkov.sirbu.06,kramkov.sirbu.07} on utility-based prices and hedging strategies for a small number of claims.}

The utility loss due to small transaction costs can also be quantified. At the leading order, it is given by the squared halfwidths of future no-trade regions, suitably averaged with respect to both time and states. Here, time is measured in business time, i.e., a clock that runs at the speed of the market's local variance: losses due to trading costs accrue more rapidly in times of frequent price moves. Averaging across states is performed under the investors' marginal pricing measure, i.e., the impact of the small costs is priced using the frictionless investors' marginal pricing rule. The key determinants for the welfare loss caused by small transaction costs are again the width of the spread, the investors' indirect risk tolerance, and the \emph{activity rate} or (squared) \emph{portfolio gamma} $d\langle \varphi \rangle_t/d\langle S \rangle_t$ of the frictionless target strategy and the market. The portfolio gamma can therefore be interpreted as a sensitivity with respect to market liquidity: passive investors with relatively inactive strategies are insensitive to changes in the spread, in stark contrast to more active traders. As observed by Rogers \cite{rogers.04}, the utility loss due to small transaction costs is composed of two parts: on the one hand, there are the direct costs incurred by actual trades. On the other hand, there is the displacement loss due to deviations from the frictionless target position. For small costs, we find that the relative sizes of these two contributions are universal, irrespective of asset and cost dynamics, and investors' preferences: transaction costs always contribute two thirds of the leading-order certainty equivalent loss, whereas the remaining one third is caused by displacement. For a small Tobin tax  \cite{tobin.78}, this implies that two thirds of the welfare lost by investors is paid out in taxes. The remaining one third dissipates due to suboptimal portfolio composition.

Our results also lead to a tractable model for trading-volume dynamics. This is one area where frictionless models fail dramatically, leading to infinite turnover on any time interval. Models with proportional transaction costs lead to finite trading volume. Yet, they also do not capture the turnover generated by a representative investor in a realistic manner, as they prescribe trading of ``bang-bang'' type: the investor either does not trade at all, or at an infinite rate. For small costs, however, we find that turnover can be approximated by a finite rate at the leading order, in line with the models typically used in the price impact literature \cite{almgren.03,garleanu.pedersen.12,guasoni.weber.12}. Implementing frictionless strategies with a constant buffer leads to trading volume proportional to the quadratic variation of the target. In contrast, the turnover generated by optimal implementation is determined by a geometric average of the local variabilities of both the frictionless target strategy and the market, scaled by risk tolerance. This increasing relation with market volatility is in line with the empirical findings of Karpoff \cite{karpoff.87}. With constant investment opportunities, constant relative risk aversion implies a constant turnover rate, whereas stochastic opportunity sets driven by stationary factors lead to stationary models for relative share turnover, allowing to reproduce empirical stylized facts such as mean reversion and autocorrelation (cf.\ Lo and Wang \cite{lo.wang.00}). At the leading order, turnover is inversely proportional to the cubic root of the spread, irrespective of the latter's future dynamics. Ceteris paribus, the model therefore predicts that reducing an already small spread by 10\% should increase turnover by about 3.6\%, irrespective of preferences as well as asset price and cost dynamics.

The above results on utility maximization extend to other widely-used optimization procedures, such as mean-variance portfolio selection in the spirit of Markowitz \cite{markowitz.52} and the Kelly criterion \cite{kelly.56} of maximizing the long-run growth rate. 

As in the frictionless case, the mean-variance optimal portfolios are obtained by rescaling the optimal strategy for (truncated) quadratic utility. Obtaining a given target return with transaction costs requires a larger multiplier and a bigger portfolio variance. Conversely, in the presence of transaction costs, a given variance bound leads to a smaller multiplier and a reduced return. Both of these effects are magnified if an ambitious target return, resp.\ loose variance bound, prescribes large positions in the risky asset. Nevertheless, the corresponding Sharpe ratio remains universal among all mean-variance optimal portfolios: it is simply decreased by a constant to account for the presence of a nontrivial spread. Here, at least two thirds of this welfare effect are caused directly by trading costs, whereas at most one third is due to displacement from the frictionless target.

In the absence of frictions, it is well known that the optimal portfolio for logarithmic utility maximizes the long-term growth rate not only in expectation, but also in an almost-sure sense. For small costs, we establish that this remains true at the leading order. As in the frictionless case, the (approximately) growth-optimal portfolio turns out to be myopic, in that it is determined completely by the local dynamics of the model. An explicit formula for the leading-order reduction of the long-run growth rate is also provided.

The remainder of the article is organized as follows. Section~\ref{sec:prelim} introduces the model and collects the inputs from the frictionless investment/consumption problem needed to formulate the leading-order corrections for small costs. The main results are presented and discussed in Section~\ref{sec:mr}. The important special case of investors with constant relative risk aversion is treated in Section~\ref{sec:crra}. Next, we turn to mean-variance portfolio selection and the growth-optimal portfolio with transaction costs, before concluding in Section \ref{sec:conclusion}. Derivations of all results are collected in Appendices \ref{sec:notation}-\ref{|sec:numeraire}. These are based on applying formal perturbation arguments to the martingale optimality conditions of a frictionless ``shadow price'' \cite{cvitanic.karatzas.96,loewenstein.00,kallsen.muhlekarbe.10}, which yields the same optimal strategy and utility as the original market with transaction costs. A rigorous verification theorem is a major challenge for future research.

\section{Preliminaries}\label{sec:prelim}

 \subsection{Setup}

Consider a financial market consisting of a safe asset with price normalized to one\footnote{As we consider general state-dependent utilities, the safe asset can be normalized without loss of generality. Indeed, for an arbitrary safe asset $S^0_t>0$, one can reduce to this case by using the latter as the numeraire and maximizing utility from discounted consumption and terminal wealth for the utilities $\widehat{u}_1(\omega,t,x)=u_1(\omega,t,x S^0_t(\omega))$ and $\widehat{u}_2(\omega,x)=u_2(\omega, xS^0_T(\omega))$.} and a risky asset, traded with small proportional transaction costs $\varepsilon_t>0$. This means purchases of the latter are carried out at a higher ask price $S_t+\varepsilon_t$, whereas sales only earn a lower bid price $S_t-\varepsilon_t$. Put differently, $\varepsilon_t$ is the halfwidth of the bid-ask spread. The mid price $S_t$ is assumed to follow a general, not necessarily Markovian, It\^o process:
$$dS_t= b^S_tdt+\sqrt{c^S_t}dW_t,$$
for a standard Brownian motion $W_t$. In this setting, an investor trades to maximize expected utility from consumption and terminal wealth,\footnote{Here, $u_1(\omega,t,x)$ and $u_2(\omega, x)$ are increasing, concave utility functions in wealth $x$, depending on time $t$ and the state $\omega$ in a nonanticipative manner.}
$$U^\varepsilon(x)=\sup_{(\psi^\varepsilon,k^\varepsilon)} E\left[\int_0^T u_1(t,k^\varepsilon_t)dt+u_2(X_T^\varepsilon(\psi^\varepsilon,k^\varepsilon))\right],$$
over all consumption rates $k_t^\varepsilon$ and trading strategies $\psi_t^\varepsilon$ with associated wealth processes\footnote{$||\phi^\varepsilon||_t$ denotes the total variation of $\phi^\varepsilon$, measuring the number of shares traded on $[0,t]$.} 
$$X_t^\varepsilon(\psi^\varepsilon,k^\varepsilon)=x+\int_0^t \psi^\varepsilon_s dS_s -\int_0^t k^\varepsilon_s ds + \Psi_t -\int_0^t \varepsilon_s d||\psi^\varepsilon||_s.$$
Here, the first two integrals describe the usual frictionless gains from trading and consumption expenditures, respectively. The third term represents the investors' cumulative endowment process, which can include both a continuous component, such as labour income, and lump-sum payments, such as an option position maturing at the terminal time $T$. Finally, the last integral accounts for the transaction costs incurred by the investors' strategy, by weighting the total variation of the latter with the width of the spread.

In most of the portfolio choice literature dating back to Constantinides and Magill \cite{magill.constantinides.76}, transaction costs equal a constant fraction of the monetary amount transacted, $\varepsilon_t= \varepsilon S_t$.  For highly liquid stocks spreads often equal a few ticks, regardless of the stock price, so that a constant spread $\varepsilon_t=\varepsilon$ may be a more plausible model. Then, transaction costs are levied on the number of shares traded as in the futures model of Jane{\v{c}}ek and Shreve \cite{janecek.shreve.10}. In general, the dynamics of the spread turn out to be inconsequential, as long as it follows an It\^o process $\mathcal{E}_t$ rescaled by a small parameter $\varepsilon$, i.e., $\varepsilon_t=\varepsilon \mathcal{E}_t$. Henceforth, $\varepsilon_t$ refers to a process of this form.

\subsection{Inputs from the Frictionless Problem}\label{ss:inputs}

In this section, we collect the inputs from the frictionless problem that determine the leading-order corrections due to the presence of small transaction costs.

Denote by $\kappa_t$, $\varphi_t$, and $X_t(\varphi,\kappa)=x+\int_0^t \varphi_s dS_s -\int_0^t \kappa_s ds+\Psi_t$ the frictionless optimal consumption rate, trading strategy, and wealth process, respectively, and write $Q$ for the corresponding marginal pricing measure.\footnote{That is, the dual martingale measure, linked to the primal optimizers by the usual first-order conditions, cf.~Appendix \ref{sec:martopt}, \cite{karatzas.zitkovic.03}, and the references therein for more details. As observed by Davis \cite{davis.97}, $Q$-expectations describe the investors' pricing rule for a marginal number of contingent claims, whence the name ``marginal pricing measure.''} 
Moreover, let $U(t,x)$ be the investors' \emph{indirect utility}, i.e., the maximal utility that can be obtained on $[t,T]$ starting from wealth $x$, by trading according to the conditionally optimal portfolio/consumption pair $(\varphi_s(t,x),\kappa_s(t,x))_{s \in [t,T]}$ (cf.\ Equation \eqref{eq:condutil} in the appendix). With this notation, define the frictionless \emph{sensitivities of consumption and investment} with respect to wealth as\footnote{Here and henceforth, primes always denote derivatives with respect to current wealth.}
$$\kappa'_t=\lim_{\delta \to 0} \frac{\kappa_t(t,X_t+\delta)-\kappa_t(t,X_t)}{\delta}, \qquad \varphi'_t=\lim_{\delta \to 0} \frac{\varphi_t(t,X_t+\delta)-\varphi_t(t,X_t)}{\delta}.$$
In the spirit of Kramkov and S{\^{\i}}rbu \cite{kramkov.sirbu.06a,kramkov.sirbu.06}, these quantities describe how marginal changes in the investors' wealth influence their consumption and investment decisions in the absence of frictions, i.e., how much of an extra dollar should be consumed or invested, respectively.

Finally, the investors' preferences are subsumed by
$$r_t=-\frac{u_1'(t,\kappa_t)}{u_{1}''(t,\kappa_t)} \quad \mbox{and} \quad R_t=-\frac{U'(t,X_t)}{U''(t,X_t)}.$$
Here, $r_t$, is the \emph{direct risk tolerance} with respect to current consumption. In contrast, $R_t$ measures the \emph{indirect risk tolerance} of the indirect utility, evaluated along the optimal frictionless wealth process.\footnote{In a Markovian setting, this is the risk tolerance of the value function, evaluated at the optimal wealth process. This object is central in the frictionless analysis of Merton \cite{merton.69,merton.71}, and also features prominently in the recent work of Soner and Touzi \cite{soner.touzi.11}.}  $R_t$ measures the investors' attitude towards future risk:
\begin{enumerate}
\item If the market is complete or the investors' preferences are described by a standard utility function of exponential or power type, then the indirect risk tolerance is given by the following conditional expectation (cf.~Section~\ref{ss:B2}):

\begin{equation}\label{eq:Rexp}
R_t=E^Q_t\left[\int_t^T -\frac{u'_1(s,\kappa_s)}{u_1''(s,\kappa_s)}ds -\frac{u'_2(X_T(\varphi,\kappa))}{u_2''(X_T(\varphi,\kappa))}\right].
\end{equation}
Whence, $R_t$ represents the investors' expected risk tolerance with respect to future consumption and terminal wealth, computed under the marginal pricing measure $Q$. 
\item Beyond complete markets and standard utility functions, \eqref{eq:Rexp} remains valid if the investors' marginal pricing measure $Q$ is replaced with some other equivalent martingale measure $\widetilde{Q}$ (see Appendix \ref{ss:B2}). Even though the latter implicitly depends on the indirect risk tolerance, one can therefore still interpret $R_t$ as an expectation of future risk tolerances with respect to consumption and terminal wealth.
\item Alternatively, the indirect risk tolerance $R_t$ can also be characterized dynamically in terms of the quadratic backward stochastic differential equation \eqref{eq:BSDE}. The sensitivities $\kappa'_t$ and $\varphi'_t$ can in turn be expressed in terms of $r_t$ and $R_t$ (cf.\ \eqref{eq:sens}).
\item If the investor focuses exclusively on utility from terminal wealth or from intermediate consumption, then the above formulas remain valid, setting $r_t=0$ or $R_T=0$, respectively. 
\end{enumerate}

 \section{Main Results}\label{sec:mr}
 
 With the inputs from the frictionless problem, the impact of small proportional transaction costs $\varepsilon_t>0$ on optimal investment and consumption policies, welfare, and implied trading volume can now be quantified as follows.

\subsection{Optimal Investment and Consumption}\label{sec:policy}

We first describe an asymptotically optimal portfolio/consumption pair. Derivations can be found in Appendix~\ref{sec:opt}.

With small transaction costs $\varepsilon_t>0$, it is approximately optimal\footnote{That is, the utility obtained from this policy is optimal at the leading order $O(\varepsilon^{2/3})$ for small costs $\varepsilon_t=\varepsilon \mathcal{E}_t$.} to consume at rate
\begin{equation}\label{eq:consumption}
\kappa^\varepsilon_t=\kappa_t +\kappa'_t\left(X_t^\varepsilon(\varphi^\varepsilon,\kappa^\varepsilon)-X_t(\varphi,\kappa)\right),
\end{equation}
while engaging in the minimal amount of trading necessary to keep the number $\varphi^\varepsilon_t$ of risky shares within the time and state dependent no-trade region $[\overline{\mathrm{NT}}_t-\Delta\mathrm{NT}_t,\overline{\mathrm{NT}}_t+\Delta\mathrm{NT}_t]$ with midpoint and halfwidth
\begin{equation}\label{eq:deviation}
\overline{\mathrm{NT}}_t=\varphi_t+\varphi'_t \left(X_t^\varepsilon(\varphi^\varepsilon,\kappa^\varepsilon)-X_t(\varphi,\kappa)\right) \quad \mbox{and} \quad 
\Delta\mathrm{NT}_t = \left(\frac{3R_t}{2} \frac{d\langle \varphi \rangle_t}{d\langle S \rangle_t} \varepsilon_t\right)^{1/3}.
\end{equation}
A first crucial observation is that the future dynamics of a stochastic spread are disregarded throughout at the leading order; only its current width is taken into account. Let us discuss the other characteristics of this policy in more detail:
\begin{enumerate}
\item Small transaction costs only influence consumption by affecting the investors' wealth. In view of \eqref{eq:consumption}, the optimal rate is simply adjusted according to the corresponding sensitivity $\kappa'_t$ of the \emph{frictionless} optimizer and the change $X_t^\varepsilon(\varphi^\varepsilon,\kappa^\varepsilon)-X_t(\varphi,\kappa)$ in wealth caused by applying the policy $(\varphi^\varepsilon_t,\kappa^\varepsilon_t)$ with frictions rather than $(\varphi_t,\kappa_t)$ without these. The sensitivity coefficient can be written as $\kappa_t'=r_t/R_t$ (cf.~\eqref{eq:sens}); hence, it is strictly positive and trades off the relative importance of present and future consumption streams. For large $r_t/R_t$, investors are less concerned about changes in their current consumption level than at later times. Hence, they are willing to deviate substantially from the frictionless target to react to changes in their wealth. If fluctuations in current consumption are deemed relatively more important than at later times, the situation is reversed. The optimal wealth with transaction costs is typically smaller than its frictionless counterpart. Therefore, small trading costs tend to reduce consumption accordingly. The greatest reductions occur at those times where investors are most tolerant with respect to changes in their consumption level. 
\item The interpretation for the midpoint $\overline{\mathrm{NT}}_t$ of the no-trade region is similar: it is shifted compared to the frictionless position $\varphi_t$ to account for the wealth effect of past transaction costs. These change the investors' optimal wealth at time $t$ from $X_t(\varphi,\kappa)$ to $X_t^\varepsilon(\varphi^\varepsilon,\kappa^\varepsilon)$, and the target position is adjusted accordingly, reminiscent of a Taylor expansion holding all other variables fixed. The scaling factor is the sensitivity $\varphi'_t=d\langle R,S \rangle_t/R_t d\langle S \rangle_t$, determined by the local dynamics of the indirect risk tolerance $R_t$ and the asset price $S_t$  (cf.~\eqref{eq:sens}). 
\item The halfwidth $\Delta\mathrm{NT}_t$ of the no-trade region is given by the cubic root of three factors.\footnote{For local utility maximizers with constant risk tolerance, an analogous result has been obtained by Martin \cite{martin.12}.} The term $\varepsilon_t$ corresponds to the absolute halfwidth of the bid-ask spread. Larger frictions require wider inactivity regions, regardless of their future dynamics. The factor $3R_t/2$ reflects the investors' tolerance to risk. Ceteris paribus, more risk tolerant investors are willing to accept larger deviations from their frictionless target in order to save transaction costs. Finally, the (squared) \emph{portfolio gamma}\footnote{If the frictionless strategy is a ``delta hedge'' in a complete Markovian setting, $\varphi_t=\Delta(t,S_t)$, then $d\langle \varphi\rangle_t/d\langle S \rangle_t=(\frac{\partial}{\partial S}\Delta(t,S_t))^2$ so that this notion indeed reduces to the square of the ``gamma'' $\Gamma(t,S_t)=\frac{\partial}{\partial S}\Delta_S(t,S_t)$.} $d\langle \varphi \rangle_t/d\langle S \rangle_t$ trades off the local activity rates of the frictionless optimal strategy and the market. Tracking highly oscillatory targets requires wide buffers to save transaction costs. Conversely, wildly fluctuating asset prices necessitate close tracking to reduce losses due to displacement from the frictionless position.
\item For utility from terminal wealth only ($u_1(t,x)=0$), the formulas for the optimal trading strategy remain valid. Consumption is of course null in this case. Conversely, the pure consumption case $(u_2(x)=0)$ is also covered by the above formulas.
\end{enumerate} 

In summary, the adjustment of the leading-order optimal policy due to small transaction costs is \emph{myopic} in the sense that it only depends on non-local quantities associated to the frictionless optimization problem, namely the investors' frictionless optimal policy and risk-tolerance wealth process. Even if the frictionless optimizer includes intertemporal hedging terms in models with stochastic opportunity sets, the effect of small trading costs is purely local: these only enter through the current width of the spread and the investors' frictional wealth.

\subsection{Welfare}\label{sec:welfare}

Now, we turn to the performance losses induced by small trading costs. Derivations can be found in Appendix~\ref{sec:uloss}.

The welfare effect of transaction costs across different models and preference structures is most easily compared in terms of certainty equivalents. To this end, let $U^\varepsilon(x)$ and $U(x)$ denote the maximal utilities that can be obtained starting from initial capital $x$,  with and without transaction costs $\varepsilon_t=\varepsilon \mathcal{E}_t$, respectively. Then, at the leading order $O(\varepsilon^{2/3})$:
\begin{equation}\label{eq:celoss}
U^\varepsilon(x) \sim U\left(x-E^Q\left[\int_0^T \frac{(\Delta\mathrm{NT}_t)^2}{2R_t}d\langle S \rangle_t\right]\right).
\end{equation}
Hence, the above $Q$-expectation represents the amount of initial capital the investor would be ready to forgo to trade the risky asset without transaction costs, i.e., the \emph{certainty equivalent loss} due to small frictions. This leading-order optimal performance is attained by the consumption/portfolio pair from Section \ref{sec:policy}. Let us discuss some of the implications of this result:
\begin{enumerate}
\item In view of \eqref{eq:celoss}, the certainty equivalent loss due to small transaction costs is determined by the future squared halfwidths of the optimal no-trade region, suitably averaged with respect to both time and states. Here, time is measured in terms of \emph{business time} $d\langle S\rangle_t$, i.e., with a clock that runs at the speed of the market's local variance. As a result, losses due to transaction costs accrue more rapidly in times of frequent price moves. Averaging across states is performed under the marginal pricing measure $Q$: the impact of small costs is priced according to the frictionless investors' marginal pricing rule.
\item The squared halfwidths of the no-trade region are normalized by the investors' risk tolerance $R_t$. The interpretation is that more risk-tolerant investors are less willing to give up initial endowment to get rid of the extra risks induced by future frictions.
\item As observed by Rogers \cite{rogers.04}, the welfare effect of transaction costs is composed of two parts. On the one hand, there are the direct costs incurred by trading. On the other hand, there is the displacement effect of having to deviate from the frictionless optimizer. At the leading order, the relative magnitudes of these two effects are universal (cf.\ Appendix \ref{sec:uloss}): two thirds of the welfare loss are caused directly by trading costs, whereas the remaining one third is due to displacement. Remarkably, this holds true irrespective of asset price and cost dynamics, as well as the investors' preference structure. For a small Tobin tax, this implies that two thirds of the corresponding certainty equivalent loss actually correspond to tax payments, whereas the remaining one third dissipates due to suboptimal portfolio composition. 
\item Fixing the investors' risk tolerance, Formulas \eqref{eq:deviation} and \eqref{eq:celoss} show that the \emph{activity rate} or \emph{portfolio gamma} $d\langle\varphi\rangle_t/d\langle S \rangle_t$ of the frictionless optimizer and the market determines the impact of a non-trivial spread $\varepsilon_t$. The (squared) portfolio gamma $d\langle \varphi \rangle_t/d\langle S \rangle_t$ therefore quantifies the investors' exposure to ``liquidity risk''.\footnote{It is important to emphasize that ``liquidity'' only refers to the width of the bid-ask spread here, and not to other proxies such as the ones proposed by, e.g., Acharya and Pedersen \cite{acharya.pedersen.05}.}  In complete markets, this notion reduces to the usual ``gamma'' of the portfolio (cf.\ the discussion in \cite{kallsen.muhlekarbe.12}), in line with the widespread interpretation of the latter as a sensitivity with respect to trading costs (see, e.g., \cite[Section 9.3]{bjoerk.03}). 
\end{enumerate}

\subsection{Implied Trading Volume Dynamics}\label{sec:impliedtv}

A severe shortcoming of frictionless diffusion models is that they lead to the absurd conclusion that the number of shares transacted  is infinite on any finite time interval. This makes it difficult to draw conclusions about the ``trading volume'' generated by a given policy.  As a remedy, one can turn to the culprit of this phenomenon, namely the Brownian component of a diffusion strategy $\phi_t$, and measure its activity in terms of its local quadratic variation $d\langle \phi \rangle_t$. But this notion of trading volume is ad hoc, and it is unclear how to relate it to the notions of \emph{share} and \emph{wealth turnover} prevalent in the empirical literature (cf., e.g., \cite{lo.wang.00}). 

Models with transaction costs present an appealing alternative, leading to finite turnover. However, a corresponding representative investor also does not match the trading volume observed in real markets: the resulting trading schemes are of ``bang-bang''-type, i.e., volume is either zero (in the no-trade region) or trades take place at an infinite rate (when the boundaries of the no-trade region are breached). Yet, as spreads decline, the trading times become more and more frequent. At the leading order, the corresponding turnover can then be approximated by a finite rate, in line with the models typically used in the price impact literature (e.g., \cite{almgren.03,garleanu.pedersen.12,guasoni.weber.12}). The resulting formulas identify the quadratic variation of the frictionless target strategy as the trading volume corresponding to suboptimal implementation with a no-trade region of constant width. In contrast, optimal rebalancing leads to a turnover rate depending on both the fluctuations of the target and the market.

To make this precise, consider a \emph{generic} frictional strategy $\phi^\varepsilon_t$ prescribing the minimal amount of trading necessary to remain inside a symmetric no-trade region $\phi_t \pm \Delta_t$ around a frictionless diffusion strategy $\phi_t$. For tight tracking ($\Delta_t \sim 0$) the corresponding \emph{absolute share turnover} is then given by (see Appendix \ref{sec:tv}):
\begin{equation}\label{eq:turnover}
||\phi^\varepsilon||_T \sim  \int_0^T \frac{d\langle \phi \rangle_t}{2\Delta_t}.
\end{equation}
At the leading order, the turnover rate is therefore determined by the ratio of the local fluctuations $d\langle \phi\rangle_t$ of the frictionless diffusion being tracked, and the width $\Delta_t$ of the no-trade region around it. Tracking a more active strategy generates higher turnover, whereas using a wider buffer decreases the required trading volume. If a buffer with constant width is used, the trading activity generated by tracking the diffusion strategy $\phi_t$ is indeed determined by the quadratic variation of the latter, up to a constant. This justifies the use of this quantity as a measure of trading activity, but only if frictionless strategies are implemented suboptimally by using a no-trade region of \emph{constant} width. For the approximately \emph{optimal} strategy $\varphi^\varepsilon_t$ from Section \ref{sec:policy}, the general formula \eqref{eq:turnover} reads as
\begin{equation}\label{eq:to}
||\varphi^\varepsilon||_T \sim \int_0^T \left(\varepsilon_t^{-1/3} \left(\frac{1}{12R_t}\right)^{1/3}\left(\frac{d\langle \varphi \rangle_t}{d\langle S \rangle_t}\right)^{2/3}\right) d\langle S\rangle_t.
\end{equation}
Measured in business time, the turnover rate corresponding to tracking the frictionless target in an approximately optimal manner is therefore determined by the cubic root of the following inputs:
\begin{enumerate}
\item The inverse of the cubic root of the absolute halfwidth $\varepsilon_t$ of the bid-ask spread. Smaller spreads allow to keep narrower buffers and therefore lead to increased turnover.
\item The investors' risk-aversion process $1/R_t$. Ceteris paribus, more risk averse investors track the frictionless target more tightly to reduce displacement losses, thereby generating higher turnover.
\item The (squared) portfolio gamma $d\langle \varphi \rangle_t /d \langle S \rangle_t$, already encountered in Sections \ref{sec:policy} and \ref{sec:welfare}; following a quickly moving target requires more adjustments.
\end{enumerate}
In calendar time, the leading-order optimal trading rate is determined by the geometric average $d\langle \varphi \rangle^{2/3}d\langle S \rangle_t^{1/3}$, scaled by risk tolerance and the spread. In particular, the turnover generated by the optimal implementation of a frictionless strategy in the presence of small transaction costs depends not only on the activity of the frictionless target, but also on the fluctuations of the market. This is in line with the empirically observed positive relationship between volume and volatility \cite{karpoff.87}. Formula \eqref{eq:to} also leads to a fully \emph{dynamic} model for turnover in the presence of a random and time-varying spread. As our model allows for random endowments, it applies to diverse types of investors, ranging from mutual funds rebalancing to maximize their long-run growth rate, high-frequency traders reacting dynamically to various signals from the market, to option desks hedging their exposure to derivative securities written on the risky asset. In each case, the resulting turnover only depends on the spread through the cubic root of its current width. All other things being equal, the model therefore predicts that reducing a small spread by 10\% should increase volume by about 3.6\%, regardless of asset and cost dynamics, preferences, and different investor types. This matches quite well with the empirical estimates of Epps \cite{epps.76}. A thorough econometric analysis of the model's testable implications is a challenging direction for future research.

\section{Constant Relative Risk Tolerance}\label{sec:crra}

The simplest special case of the above general results is given by investors with exponential utilities. Their \emph{absolute} risk tolerances with respect to intermediate consumption and terminal wealth are constant, leading to a deterministic indirect risk tolerance even with a random endowment stream (see Appendix~\ref{ss:B2}). This setting and applications to utility-based pricing and hedging are discussed at length in the companion paper of the present study \cite{kallsen.muhlekarbe.12}. 

In this section, we specialize the general results of the previous section to the specification most widely used in the literature on portfolio choice, namely \emph{isoelastic utilities} with constant \emph{relative} risk tolerance $1/\gamma>0$. More specifically, fix an impatience rate $\delta>0$, a scaling factor $\beta \geq 0$, and suppose the utilities from consumption and terminal wealth are both of power type, i.e., $u_1(t,x)=\beta e^{\delta(T-t)}x^{1-\gamma}/(1-\gamma)$ and $u_2(x)= x^{1-\gamma}/(1-\gamma)$, resp.\  $u_1(t,x)=\beta e^{\delta(T-t)}\log x$ and $u_2(x)= \log x$ for $\gamma=1$.
Without random endowments ($\Psi_t=0$), constant relative risk tolerance implies that the investors' indirect risk tolerances are given by a constant multiple of their optimal wealth processes (cf.\ Appendix~\ref{ss:B2}), $R_t=X_t(\varphi,\kappa)/\gamma$. As in the frictionless case, all quantities of interest are therefore most conveniently expressed in relative terms.\footnote{As is well known, random endowments generally destroy the homotheticity of the problem and therefore rule out the simplifications offered by this change of parametrization.}  To wit, write the dynamics of the risky asset in terms of returns, $dS_t/S_t=dY_t$, and express portfolio/consumption pairs in terms of the risky weight\footnote{At the leading order, it does not matter whether this fraction is evaluated at the bid-, ask-, mid-, or any other price process taking values in the bid-ask spread.} $\pi^\varepsilon_t=\varphi^\varepsilon_t S_t/X^\varepsilon_t(\varphi^\varepsilon,\kappa^\varepsilon)$ and the consumption/wealth ratio $c^\varepsilon_t=\kappa^\varepsilon_t/X^\varepsilon_t(\varphi^\varepsilon,\kappa^\varepsilon)$. Finally, parametrize transaction costs as fractions of current stock prices, $\varepsilon_t=\eta_t S_t$. With these notions, the wealth effect of past transaction costs is absorbed, so that  (\ref{eq:consumption}--\ref{eq:deviation}), \eqref{eq:celoss}, and \eqref{eq:to} directly lead to particularly simple formulas for the optimal policy as well as the associated welfare and trading volume.

\subsection{Optimal Investment and Consumption}\label{sec:investpower}
For investors with constant relative risk tolerance, it follows from (\ref{eq:consumption}--\ref{eq:deviation}) that it is approximately optimal to consume the same fraction of current wealth as in the frictionless case, $c^\varepsilon_t=c_t$, and engage in the minimal amount of trading necessary to keep the risky fraction $\pi^\varepsilon_t$ within a symmetric no-trade region $[\pi_t-\Delta\pi_t,\pi_t+\Delta\pi_t]$ around the frictionless target weight $\pi_t$. In view of \eqref{eq:deviation}, the halfwidth $\Delta\pi_t$ is given explicitly in terms of the local dynamics of $\pi_t$ and the  return process $dY_t=dS_t/S_t$:
\begin{equation}\label{eq:boundariespower}
\Delta\pi_t =\left(\frac{3\eta_t}{2\gamma} \left( \pi_t^2(1-\pi_t)^2-2\pi_t(1-\pi_t)\frac{d\langle \pi,Y\rangle_t}{d\langle Y \rangle_t}+\frac{d\langle \pi \rangle_t}{d\langle Y \rangle_t}\right)\right)^{1/3}.
\end{equation}
Several remarks are in order here:
\begin{enumerate}
\item The investors' optimal consumption/wealth ratio is unaffected by the presence of small frictions. Whereas the adverse effect of the transaction costs tends to reduce the absolute level of consumption, the fraction of current wealth to be consumed remains the same. 
\item Up to the factor $3\eta_t/2\gamma$, which only depends on the relative spread $\eta_t$ and the investors' preferences but not on the asset dynamics, the half-width of the no-trade region is completely determined by the frictionless optimal risky fraction $\pi_t$ as well as the joint dynamics of the latter and the return process $Y_t$. In models that can be solved in closed-form in the absence of frictions (e.g., \cite{merton.69,merton.71, kim.omberg.96,wachter.02,chacko.viceira.05,liu.07}), this immediately yields explicit formulas for the optimal no-trade regions with small costs. 

 The first term, $\pi_t^2(1-\pi_t)^2$, already arises in the  Black-Scholes model \cite{janecek.shreve.04,bichuch.11,gerhold.al.11}. It corresponds to the buffer necessary to account for changes in the risky fraction due to price moves of the risky asset. Accordingly, this term vanishes for $\pi_t=0$ or $\pi_t=1$, when full investment in either the safe or the risky asset locally immunizes the investors' risky fraction from price changes of the risky asset.

 The last term, $d\langle \pi\rangle_t/d\langle Y \rangle_t$, comes into play in models with stochastic opportunity sets, where the frictionless target weight $\pi_t$ is typically no longer constant. Here, the interpretation is similar to the boundaries expressed in numbers of risky shares: there is a tradeoff between the local fluctuations of the target and the returns of the risky asset. Wildly fluctuating targets require a wide buffer to save transaction costs, however, quickly oscillating asset prices require close tracking to limit displacement from the frictionless portfolio composition. 
 
The first and the last term are always positive. The second term $-2\pi_t(1-\pi_t)d\langle \pi,Y\rangle_t/d\langle Y\rangle_t$, however, can be either positive or negative, depending both on whether the investors' position is leveraged ($\pi_t>1$) or not, and on the correlation between shocks to returns and the frictionless weight. The interpretation is that the no-trade region can be narrowed if the fluctuations of the frictionless number of shares is reduced because shocks to returns and the target weight partially offset. Conversely, the no-trade region is widened if the directions of the two effects tend to agree. Accordingly, the sign of this term switches when passing from an unlevered to a levered position, because positive returns then lead to negative shocks to the risky weight.
 
 \item In the special case of logarithmic utility, the frictionless optimal portfolio/consumption policy can be determined explicitly in general \cite{merton.71}. The optimal consumption/wealth ratio is then completely determined by the investors' time horizon and their impatience rate, whereas the corresponding risky weight $\pi_t$ should be held equal to the Merton proportion, i.e.,  the market's (infinitesimal) mean-variance ratio. Put differently, the problem is \emph{myopic}, i.e., completely determined by the local dynamics of the underlying asset price model, and consumption and investment separate. The results above show that both of these findings are robust with respect to small frictions: the optimal consumption/wealth ratio remains the same, and the trading boundaries are completely determined by the local dynamics of the model, just like the optimal frictionless weight. 
 \item The formulas above  remain valid if the investor only focuses on utility from intermediate consumption or from terminal wealth, setting consumption to zero in the second case.
\end{enumerate}

\subsection{Welfare}
To conveniently express the certainty equivalent loss \eqref{eq:celoss} due to small transaction costs in terms of relative quantities, denote by $\widehat{P}$ the measure with density process $\E(\int_0^\cdot \pi_s dY_s)_t$ relative to the marginal pricing measure $Q$.\footnote{In the absence of consumption, $\E(\int_0^\cdot \pi_s dY_s)_t$ is the optimal frictionless wealth process starting from unit capital.} Then, the leading-order certainty equivalent loss due to small transaction costs can be described as follows:
\begin{equation}\label{eq:utilityloss}
U^\varepsilon(x) \sim U\left(x\left(1-E^{\widehat{P}}\left[\frac{\gamma}{2}\int_0^T \frac{(\Delta\pi_t)^2}{\E(\int_0^\cdot c_u du)_t} d\langle Y \rangle_t \right]\right)\right).
\end{equation}
Hence, the above $\widehat{P}$-expectation is the fraction of initial capital the investor would be willing to give up to trade the risky asset without transaction costs. This is the scale-invariant measure for the effect of transaction costs computed numerically by Balduzzi and Lynch \cite{balduzzi.lynch.99}.
 \begin{enumerate}
\item Without intermediate consumption ($c_t=0$), the above formula for the certainty equivalent loss can also be interpreted in term of \emph{equivalent safe rates} as in \cite{gerhold.al.11}. To wit, let $\rho_T, \rho^\varepsilon_T$ be fictitious safe rates, for which a full safe investment yields the same utility as trading optimally in the original market without and with transaction costs, respectively:
$$U(x)=u_2(xe^{\rho_T T}) \quad \mbox{resp.} \quad  U^\varepsilon(x)=u_2(xe^{\rho^\varepsilon_T T}).$$
In view of the homotheticity $U(x)=x^{1-\gamma}U(1)$ of the indirect utility inherited from the isoelastic utility $u_2(x)=x^{1-\gamma}/(1-\gamma)$, this gives
$$\rho_T=\frac{1}{T}\log u_2^{-1}(U(1)) \quad \mbox{resp.}\quad  \rho^\varepsilon_T=\frac{1}{T}\log u_2^{-1}(U^\varepsilon(1)).$$
Therefore, also taking into account $u_2^{-1}(x)=((1-\gamma)x)^{1/(1-\gamma)}$, \eqref{eq:utilityloss}, and Taylor expanding:
\begin{align}
\rho^\varepsilon_T=\rho_T + \frac{1}{(1-\gamma)T} \log\frac{U^\varepsilon(1)}{U(1)} &\sim \rho_T + \frac{1}{(1-\gamma)T} \log\frac{U\left(1-E^{\widehat{P}}\left[\frac{\gamma}{2}\int_0^T (\Delta\pi_t)^2 d\langle Y \rangle_t \right]\right)}{U(1)} \notag \\
&\sim \rho_T - \frac{1}{T} E^{\widehat{P}}\left[\frac{\gamma}{2}\int_0^T (\Delta\pi_t)^2 d\langle Y \rangle_t \right]. \label{eq:esr}
\end{align}
Hence, without intermediate consumption, the certainty equivalent loss per year and unit of initial wealth also admits an alternative interpretation as the reduction of the equivalent safe rate due to small transaction costs.
\item In the absence of consumption ($c_t=0$), the impact of small transaction costs is, up to a constant, given by the squared-halfwidth $(\Delta\pi_t)^2$ of the no-trade region, suitably averaged with respect to both time and states. Here, time is measured in terms of $d\langle Y \rangle_t$, i.e., the activity of the returns. Averaging across states is performed under the auxiliary measure $\widehat{P}$ which, incidentally, also appears in the asymptotic analysis of utility-based prices and hedging strategies by Kramkov and S{\^{\i}}rbu \cite{kramkov.sirbu.06,kramkov.sirbu.07}. Moreover, it coincides with the \emph{myopic probability} of Guasoni and Robertson \cite{guasoni.robertson.12}, under which a hypothetical log-investor chooses the same optimal policy as the original isoelastic investor under the physical probability. Accordingly, $\widehat{P}=P$ if the original isoelastic investor has a logarithmic utility with unit risk tolerance.
\item With a nontrivial consumption/wealth ratio $c_t>0$, the future half-widths $(\Delta\pi_t)^2$ of the no-trade region are ``discounted'' by $\E(\int_0^\cdot c_s ds)_t=\exp(\int_0^t c_s ds)$, a measure of consumption accrued until then. This takes into account that intermediate consumption reduces wealth, so that the same trading boundaries in terms of fractions of wealth lead to less turnover and thereby a smaller welfare effect of transaction costs.
\end{enumerate}

\subsection{Implied Trading Volume Dynamics}

Just like trading policies and welfare, turnover is best expressed in terms of relative quantities for power utilities. Then, \eqref{eq:to} immediately leads to tractable formulas for the measures typically used in the empirical literature \cite{lo.wang.00}. More specifically, \emph{relative share turnover} (number of shares traded divided by number of shares held) is given by
\begin{align*}
\mathrm{ShTu}_T&=\int_0^T \frac{d||\varphi^\varepsilon||_t}{|\varphi^\varepsilon_t|} \\
&\sim \int_0^T \left(\frac{12\eta_t}{\gamma}\right)^{-1/3} \frac{1}{|\pi_t|} \left(\pi_t^2(1-\pi_t)^2 -2\pi_t (1-\pi_t) \frac{d\langle \pi, Y \rangle_t}{d\langle Y \rangle_t}+\frac{d\langle \pi \rangle_t}{d\langle Y \rangle_t}\right)^{2/3}d\langle Y \rangle_t,
\end{align*}
at the leading order. Similarly, \emph{relative wealth turnover} (wealth transacted divided by wealth held) can be expressed as
\begin{align*}
\mathrm{WeTu}_T&=\int_0^T \frac{S_t d||\varphi^\varepsilon||_t}{X_t^\varepsilon(\varphi,\kappa)}\\
& \sim  \int_0^T \left(\frac{12\eta_t}{\gamma}\right)^{-1/3} \left(\pi_t^2(1-\pi_t)^2 -2\pi_t (1-\pi_t) \frac{d\langle \pi, Y \rangle_t}{d\langle Y \rangle_t}+\frac{d\langle \pi \rangle_t}{d\langle Y \rangle_t}\right)^{2/3}d \langle Y \rangle_t.
\end{align*}
If the frictionless target $\pi_t$ is constant, these formulas reduce to the constant rates of Gerhold et al.~\cite{gerhold.al.11}. Generally, they are \emph{stationary} in that their time averages converge to a long-run mean if the transaction costs $\eta_t$, the frictionless risky weight $\pi_t$, and the market's variance $d\langle Y\rangle_t/dt$ are all driven by stationary factors. Moreover, the turnover rate inherits properties such as mean-reversion and autocorrelation from the model's state variables, which is consistent with findings from the empirical literature \cite{lo.wang.00}.

\section{Mean-Variance Portfolio Selection}\label{ref:resultmvps}

We now turn to \emph{mean-variance portfolio selection} with transaction costs; the results reported here are derived in Appendix~\ref{sec:mvps}. Dating back to the seminal work of Markowitz \cite{markowitz.52}, Tobin \cite{tobin.58} and Merton \cite{merton.72}, mean-variance criteria have been widely used, both in theory and practice. Here, a portfolio is called \emph{mean-variance optimal} or \emph{efficient} if it minimizes the variance for a given mean or, equivalently, maximizes the mean for a given variance bound. Fix an initial endowment $x$ and a time horizon $T>0$. Then, as recapitulated in Appendix~\ref{ss:mvps1} (also cf.\ \cite{richardson.89,schweizer.94,zhou.li.00,cerny.kallsen.07}), the optimal portfolio is obtained from the one for the quadratic utility function $u(x)=-x^2$ as follows. Let $U(-1)$ and $\phi_t$ be the maximal expected quadratic utility and the corresponding optimal portfolio for the standardized initial endowment $-1$. To achieve a target mean $E[X_T(\varphi)]=m>x$ with minimal variance, trade the $(m-x)/(1+U(-1))$-fold  $\varphi_t$ of the optimal strategy $\phi_t$ for quadratic utility, and keep the remaining initial endowment $x+(m-x)/(1+U(-1))$ invested in the safe asset. The corresponding minimal variance is given by
\begin{align*}
\mathrm{Var}[X_T(\varphi)]=(m-x)^2 \frac{-U(-1)}{1+U(-1)}.
\end{align*}
In particular, this leads to the well-known result that the Sharpe ratio is the same for any mean-variance efficient portfolio, irrespective of the target mean $m>x$:
$$\mathrm{SR}= \frac{E[X_T(\varphi)]-x}{\sqrt{\mathrm{Var}[X_T(\varphi)]}}=\sqrt{-\frac{1}{U(-1)}-1}.$$
This also determines the maximal return for a given variance bound $s^2$ as
$$E[X_T(\varphi)-x]=s\mathrm{SR}.$$

Let us now discuss how these results adapt to the presence of small transaction costs.\footnote{In the Black-Scholes model, transaction costs of arbitrary size have been  considered by Dai, Xu, and Zhou \cite{dai.al.10}.} As derived in Appendix \ref{ss:mvps2}, the optimal portfolio is again obtained by rescaling its counterpart for quadratic utility. To wit, denoting the latter by $\phi^\varepsilon_t$ and writing $U^\varepsilon(-1)$ for the corresponding maximal expected quadratic utility for initial endowment $-1$, the mean-variance efficient strategy $\varphi^\varepsilon_t$ with target return $E[X^\varepsilon_T(\varphi)]=m>x$ is given by the $(m-x)/(1+U^\varepsilon(-1))$-fold of $\phi^\varepsilon_t$, where the remaining initial endowment $x+(m-x)/(1+U^\varepsilon(-1))$ is again held in the safe asset. By applying the general results of Section \ref{sec:mr} to (truncated) quadratic utility (see Appendix \ref{ss:mvps2} for more details), this shows that the mean-variance optimal strategy for a given target mean $m>x$ and small transaction costs $\varepsilon_t$ is to keep the number of risky shares in the no-trade region $[\overline{\mathrm{NT}}-\Delta\mathrm{NT},\overline{\mathrm{NT}}+\Delta\mathrm{NT}]$, with half-width $\Delta\mathrm{NT}_t$ and midpoint $\overline{\mathrm{NT}}_t$ obtained from their counterparts $\Delta\mathrm{NT}^\phi_t$ and $\overline{\mathrm{NT}}^\phi_t$ for the optimal portfolio $\phi^\varepsilon_t$ with quadratic utility and initial endowment $-1$ by rescaling:
\begin{gather*}
\Delta\mathrm{NT}_t=\frac{m-x}{1+U^\varepsilon(-1)} \Delta\mathrm{NT}^\phi_t, \quad \mbox{where } \Delta\mathrm{NT}^\phi_t=\left(\frac{3\varepsilon_t(1-\int_0^t \phi_s  dS_s)}{2}  \frac{d\langle\phi \rangle_t}{d\langle S\rangle_t}\right)^{1/3},\\
\overline{\mathrm{NT}}_t = \frac{m-x}{1+U^\varepsilon(-1)} \overline{\mathrm{NT}}^\phi_t, \quad \mbox{where } \overline{\mathrm{NT}}^\phi_t= \phi_t \frac{X^\varepsilon_t(\phi^\varepsilon)}{X_t(\phi)}.
\end{gather*}
Compared to the frictionless case, the multiplier is increased from $(m-x)/(1+U(-1))$ to
$$\frac{m-x}{1+U^\varepsilon(-1)} \sim \frac{m-x}{1+U(-1)}\left(1+2\frac{-U(-1)}{1+U(-1)}E^Q\left[\int_0^T \frac{(\Delta\mathrm{NT}^\phi_t)^2}{2(1-\int_0^t \phi_s dS_s)}d\langle S \rangle_t\right]\right),$$ 
where $Q$ denotes the variance-optimal martingale measure with density $dQ/dP=X_T(\phi)/U(-1)$. Hence, larger frictions require bigger multipliers to achieve the same target return. The corresponding minimal variance required to obtain the same return $E[X_T^\varepsilon(\varphi^\varepsilon)]=m>x$ with transaction costs is increased to 
\begin{align*}
\mathrm{Var}[X^\varepsilon_T(\varphi^\varepsilon)]&=(m-x)^2\frac{-U^\varepsilon(-1)}{1+U^\varepsilon(-1)}\\
&\sim \mathrm{Var}[X_T(\varphi)]\left(1+\frac{2}{1+U(-1)}E^Q\left[\int_0^T \frac{(\Delta\mathrm{NT}^\phi_t)^2}{2(1-\int_0^t \phi_s dS_s)}d\langle S \rangle_t\right]\right).
\end{align*}
Conversely, small transaction costs reduce the maximal expected return for a given variance bound $s^2$ from $s \mathrm{SR}$ to
$$E[X_T^\varepsilon(\varphi^\varepsilon)-x]=s\mathrm{SR}^\varepsilon \sim s \left(\mathrm{SR} -\frac{1+\mathrm{SR}^2}{\mathrm{SR}} E^Q\left[\int_0^T \frac{(\Delta\mathrm{NT}^\phi_t)^2}{2(1-\int_0^t \phi_s dS_s)}d\langle S \rangle_t\right]\right).$$
The portfolio adjustments as well as the increases in variance resp.\ decreases in returns are bigger for more ambitious target returns resp.\ looser variance bounds. Nevertheless, the maximal Sharpe ratios remain universal among mean-variance efficient portfolios also with small transaction costs:
$$\mathrm{SR}^\varepsilon  \sim \mathrm{SR}-\frac{1+\mathrm{SR}^2}{\mathrm{SR}} E^Q\left[\int_0^T \frac{(\Delta\mathrm{NT}^\phi_t)^2}{2(1-\int_0^t \phi_s dS_s)}d\langle S \rangle_t\right].$$

In summary, mean-variance optimal strategies are obtained by rescaling the optimal portfolio for (truncated) quadratic utility like in the absence of frictions. To obtain the same target return (resp.\ variance bound) even with transaction costs, the multiplier and the corresponding variance have to be increased (resp.\ the correponding return has to be decreased). Even though this leads to wider no-trade regions for higher target returns (resp.\ larger variance bounds), the Sharpe ratio is still the same for all mean-variance efficient portfolios: it is simply reduced by a constant to account for the presence of a non-trivial bid-ask spread. The reduction of the optimal Sharpe ratio due to small transaction costs can again be decomposed into direct trading costs, and displacement from the frictionless target. Here, transaction costs directly contribute at least two thirds of the Sharpe ratio loss, whereas at most one third is caused by displacement.

\section{The Growth-Optimal Portfolio} 

Consider the model without random endowment stream ($\Psi_t=0$). Then, as already pointed out in Section \ref{sec:investpower} above, the optimal trading boundaries \eqref{eq:boundariespower} for logarithmic utility do not depend on the time horizon, as they are completely determined by the local dynamics of the asset prices and the myopic frictionless optimizer. Without intermediate consumption, this policy in turn maximizes logarithmic utility $E[\log X_T^\varepsilon(\psi^\varepsilon)]$ on \emph{any} horizon $T>0$, and thereby also the \emph{expected growth rate of wealth} $\limsup_{T \to \infty} \frac{1}{T} E[\log X^\varepsilon_T(\psi^\varepsilon)]$, both at the leading order for small costs. 

As in the frictionless case \cite[Theorem 3.10.1]{karatzas.shreve.98}, this property can be strengthened in an almost sure sense. Indeed, the log-optimal portfolio $\varphi^\varepsilon_t$ maximizes the \emph{almost sure long-term growth rate}, at the leading order for small costs:
\begin{equation}\label{eq:growth}
\limsup_{T \to \infty} \frac{1}{T} \log X_T^\varepsilon(\psi^\varepsilon) \leq \limsup_{T \to \infty} \frac{1}{T} \log X_T^\varepsilon(\varphi^\varepsilon) +o(\varepsilon^{2/3}),
\end{equation}
for all competitors $\psi^\varepsilon$. In the frictionless case, this property has attracted the attention of various authors dating back to Kelly \cite{kelly.56}, Latan\'e \cite{latane.59}, Markowitz \cite{markowitz.59}, and Breiman \cite{breiman.60}. The trading boundaries \eqref{eq:boundariespower} provide the corresponding leading-order correction for small proportional transaction costs, characterized explicitly in terms of the local asset dynamics as in the fricitionless case. The impact of a small spread on the maximal asymptotic growth rate can also be quantified explicitly. At the leading order, it is given by
\begin{equation}\label{eq:growthrate}
\lim_{T \to \infty} \frac{1}{T} \log X_T(\varphi)-\lim_{T \to \infty} \frac{1}{T} \log X_T^\varepsilon(\varphi^\varepsilon)  = \lim_{T \to \infty} \frac{1}{T} \int_0^T \frac{(\Delta\pi_t)^2}{2} d\langle Y \rangle_t,
\end{equation}
if the limits exist. Hence, the reduction of the long-term growth rate due to small transaction costs is given by the long-run average squared halfwidth of the no-trade region, measured in risky fractions and computed in business time. For concrete models with stationary state variables, this readily yields explicit formulas by integrating against the corresponding invariant measures. The derivation of these results can be found in Appendix~\ref{|sec:numeraire}.

\section{Conclusion}\label{sec:conclusion}

This paper provides simple asymptotic formulas for optimal investment and consumption policies as well as the corresponding welfare and implied trading volume, in the practically relevant limiting regime of small bid-ask spreads. 

We find that investors should keep their positions in a time- and state-dependent no-trade region, which is myopic in that it is fully determined by the current spread, the investors' frictional wealth, and quantities inferred from the frictionless version of the problem.  The corresponding welfare effect of small transaction costs is determined by the squared halfwidths of future no-trade regions, suitably averaged across time and states. Here, the investors' strategy enters through the activity rate $d\langle \varphi \rangle_t/d\langle S \rangle_t$ -- the portfolio's squared gamma in complete markets -- which therefore quantifies the exposure to liquidity risk through changes in the spread. At the optimum, two thirds of the utility loss are incurred due to actual trading costs, whereas the remaining one third is caused by displacement from the frictionless target. Optimal implementation of frictionless strategies in the presence of a small bid-ask spread leads to a turnover rate determined by the local fluctuations of both the target strategy and the target; the spread only enters though the cubic root of its current width.

All of these results are surprisingly robust: they apply for general diffusive asset price and cost dynamics, arbitrary preferences over intermediate consumption and terminal wealth, and also in the presence of random endowment streams.  Moreover, they can be extended to cover other widely-used optimization criteria such as mean-variance portfolio selection and the maximization of the long-term growth rate.

\appendix
\section*{Appendix: Derivation of the Main Results}

In the sequel, we derive the results presented in the previous sections. Throughout, mathematical formalism is treated liberally. For example, we do not state and
verify technical conditions warranting the uniform integrability of local martingales, interchange of integration and differentiation, and the uniformity of estimates. In particular, the Landau symbols $O(\cdot)$ and $o(\cdot)$ refer to pointwise estimates, with the implicit assumption of enough regularity in time and states to eventually turn these into an estimate of the expected utility generated by the approximately optimal policy. Full verification theorems have been worked out by Soner and Touzi~\cite{soner.touzi.11} as well as Possama\"i, Soner and Touzi \cite{possamai.al.12} for infinite-horizon consumption problems in complete Markovian markets. 

\section{Notation}\label{sec:notation}
Throughout, we write $\phi \mal S_t$ for the stochastic integral $\int_0^t \phi_s dS_s$ and denote the identity process by $I_t = t$. For an It\^o process $X_t$, we write $b^X_t$ for its drift rate, resp.\ $b_t^{X,Q}$ if the latter is computed under another measure $Q$. Moreover, we denote by $c^X_t=d\langle X \rangle_t/dt$ its local quadratic variation, and by $c^{XY}_t=d\langle X,Y \rangle_t/dt$ its local covariation with another It\^o process $Y_t$.

\section{The Risk-Tolerance Process}\label{sec:risktol}

We begin with an analysis of the investors' indirect risk tolerance introduced in Section~\ref{ss:inputs}, which describes how the frictionless investors' attitude towards future risks changes with their wealth level. It thereby determines how the wealth effect of a small friction is reflected in the optimal policy and utility. As a result, we expect it to play a pivotal role not only for the proportional transaction costs considered here but also in the asymptotic analysis of other market imperfections.

To define the investors' indirect risk tolerance, denote by $X_t=X_t(\varphi,\kappa)$ the optimal wealth process of the frictionless utility maximization problem, generated by the optimal portfolio/consumption pair $(\varphi_t,\kappa_t)$. Define the investors' \emph{indirect utility} as
\begin{equation}\label{eq:condutil}
U(t,x)=\sup_{(\psi_s,k_s)_{s\in [t,T]}} E_t\left[\int_t^T u_1(s,k_s)ds+u_2\left(x+\int_t^T \psi_s dS_s -\int_t^T k_s ds + \Psi_t\right)\right],
\end{equation}
where the supremum is taken over all portfolio/consumption pairs on $[t,T]$. Then, we call the risk tolerance $R_t=-U'(t,X_t)/U''(t,X_t)$ of the indirect utility, evaluated along the optimal wealth process, \emph{indirect risk tolerance}. In this section, we investigate the properties of this object. First, we describe its local dynamics in terms of a quadratic backward stochastic differential equation (henceforth BSDE). Then, we discuss how it can be represented as a suitable expectation of the terminal risk tolerance $-u_2'(X_T)/u_2''(X_T)$ and the intermediate risk tolerances $-u_1'(t,\kappa_t)/u_1''(t,\kappa_t)$ from consumption, and describe how it generalizes the \emph{risk-tolerance wealth process} of Kramkov and S{\^{\i}}rbu \cite{kramkov.sirbu.06}.

\subsection{A Dynamic Characterization of the Indirect Risk Tolerance}

Our starting point is the dynamic programming principle, which states that, for any infinitesimal interval $dt$:
\begin{align}
U(t-dt,x)&=\sup_{(\psi_t,k_{t-dt})}\Big(u_1(t-dt,k_{t-dt})dt+E_{t-dt}[U(t,x+\psi_t dS_t- k_{t-dt} dt)]\Big)\notag\\
&=:\sup_{(\psi_t,k_{t-dt})} f(t-dt,x,\psi_t,k_{t-dt}) =:f(t-dt,x,\varphi_t(x),\kappa_{t-dt}(x)).\label{eq:deff}
\end{align}
Here, $\varphi_t(x)$ and $\kappa_{t-dt}(x)$ denote the (predictable) optimal number of risky shares and the (adapted) optimal consumption given wealth $x$ at time $t-dt$. Evaluated along the optimal wealth process, these coincide with the globally optimal portfolio/consumption pair $\varphi_t,\kappa_{t-dt}$. The optimality of $\varphi_t(x),\kappa_{t-dt}(x)$ implies that the respective partial derivatives vanish:
\begin{equation}\label{eq:zero}
f_{\varphi}(t-dt,x,\varphi_t(x),\kappa_{t-dt}(x))=0, \qquad f_\kappa(t-dt,x,\varphi_t(x),\kappa_{t-dt}(x))=0.
\end{equation}
Moreover, differentiating the function $f(t-dt,x,\varphi_t(x),\kappa_{t-dt}(x))$ defined in \eqref{eq:deff} with respect to $\varphi_t(x)$ and $\kappa_{t-dt}(x)$, we obtain
\begin{align}
f_{\varphi\kappa}(t-dt,x,\varphi_t(x),\kappa_{t-dt}(x))&=\partial_\kappa E_{t-dt}[U'(t,x,\varphi_t(x)dS_t-\kappa_{t-dt}(x)dt)dS_t]\nonumber\\
&=E_{t-dt}[U''(t,x,\varphi_t(x)dS_t-\kappa_{t-dt}(x)dt)dS_t dt]=0\label{eq:mixed}.
\end{align}
In the next calculations, we suppress the arguments of the functions to ease notation. In view of \eqref{eq:zero}:
\begin{align}\label{eq:ux}
U'&=f_x+f_{\varphi}\varphi_t'(x)+f_\kappa\kappa'_{t-dt}(x)=f_x.
\end{align}
Moreover, \eqref{eq:zero} and \eqref{eq:mixed} also yield
\begin{align*}
0=\frac{\mathrm{d}0}{\mathrm{d}x}&= \frac{\mathrm{d} f_{\varphi}}{\mathrm{d} x}= f_{\varphi x}+f_{\varphi\varphi} \varphi'_t(x), \qquad 0=\frac{\mathrm{d} f_\kappa}{\mathrm{d} x}=f_{\kappa x}+f_{\kappa\kappa}\kappa'_{t-dt}(x),
\end{align*}
and therefore\footnote{Note that differentiation of the function $f$ from \eqref{eq:deff} shows that both denominators are strictly positive.}
\begin{equation}\label{eq:sensitivity}
\varphi_t'(x)=-\frac{f_{\varphi x}}{f_{\varphi\varphi}},  \qquad	\kappa_{t-dt}'(x) = -\frac{f_{\kappa x}}{f_{\kappa\kappa}}.
\end{equation}
As a result:
\begin{align}
U''&=f_{xx}+f_{x\varphi}\varphi'_t(x)+f_{x\kappa}\kappa_{t-dt}'(x)=f_{xx}-\frac{(f_{x\varphi})^2}{f_{\varphi\varphi}}-\frac{(f_{x\kappa})^2}{f_{\kappa\kappa}}.		\label{eq:uxx}
\end{align}

With these preparations, we now argue by formal recursion that the indirect marginal utility $U'(t,X_t)$ evaluated along the optimal wealth process coincides with the dual martingale density $Z_t$, which is -- up to normalization -- the density process of the marginal pricing measure $Q$. This is evidently true at the terminal time $T$, where $U'(T,X_T)=u'_2(X_T)=Z_T$ reduces to the well-known first-order condition linking the solutions of the primal and dual problems (see, e.g., \cite[Section 3.6]{karatzas.shreve.98}). Now, suppose that it is already known that 
\begin{equation}\label{eq:hyp}
U'(t,X_{t})=Z_{t}.
\end{equation}
Then, \eqref{eq:ux} and the definition of $f$ give
$$U'(t-dt,x)=f_x(t-dt,x,\varphi_t(x),\kappa_{t-dt}(x))=E_{t-dt}[U'(t,x+\varphi_t(x)dS_t-\kappa_{t-dt}(x)dt)].$$
Evaluated at the optimal wealth $X_{t-dt}$ and using \eqref{eq:hyp}, this shows
\begin{equation}\label{eq:z}
U'(t-dt,X_{t-dt})=E_{t-dt}[Z_{t}]=Z_{t-dt}.
\end{equation}
Here, the last equality follows from the martingale property of $Z_t$, completing the recursion. Also note that, together with \eqref{eq:zero}, this gives 
$$0=f_\kappa(t-dt,X_{t-dt},\varphi_t,\kappa_{t-dt})=u'_1(t-dt,\kappa_{t-dt})dt+E_{t-dt}[-Z_tdt]=u'_1(t-dt,\kappa_{t-dt})dt-Z_{t-dt}dt,$$
and thereby 
\begin{equation}\label{eq:Zu1}
Z_{t-dt}=u'_1(t-dt,\kappa_{t-dt}).
\end{equation}

Next, we turn to the indirect risk tolerance $R_t=-U'(t,X_t)/U''(t,X_t)$, and describe its dynamics by means of a BSDE. To achieve this, it is easier to start from the indirect risk aversion $R_t^{-1}=-U''(t,X_t)/U'(t,X_t)=-U''(t,X_t)/Z_t$. In view of \eqref{eq:uxx}, we first compute the various partial derivatives of $f$. By definition of $f$, we have
\begin{align*}
&f_{xx}(t-dt,x,\varphi_t(x),\kappa_{t-dt}(x))\\
&\qquad =E_{t-dt}[U''(t,x+\varphi_t(x)dS_t-\kappa_{t-dt}(x)dt)]\\
&\qquad =E_{t-dt}\left[U'(t,x+\varphi_t(x)dS_t-\kappa_{t-dt}(x)dt) \frac{U''(t,x+\varphi_t(x)dS_t-\kappa_{t-dt}(x)dt)}{U'(t,x+\varphi_t(x)dS_t-\kappa_{t-dt}(x)dt)}\right].
\end{align*}
Evaluating at $X_{t-dt}$, inserting the definition of $R_t^{-1}$ as well as \eqref{eq:hyp}, and using the generalized Bayes' rule, it follows that
\begin{align}
f_{xx}(t-dt,X_{t-dt},\varphi_t,\kappa_{t-dt})=-Z_{t-dt} E^Q_{t-dt}[R^{-1}_{t}] &=-Z_{t-dt} E^Q_{t-dt}[R^{-1}_{t-dt}+dR^{-1}_t]\notag\\
&=-Z_{t-dt}(R_{t-dt}^{-1}+b_t^{R^{-1},Q}dt).\label{eq:1}
\end{align}
Next, 
$$f_{x\varphi}(t-dt,x,\varphi_t(x),\kappa_{t-dt}(x))=E_{t-dt}[U''(t,x+\varphi_t(x)dS_t-\kappa_{t-dt}(x)dt)dS_t].$$
Similarly as above, evaluation at $X_{t-dt}$ gives
\begin{align}\label{eq:2}
f_{x\varphi}(t-dt,X_{t-dt},\varphi_t,\kappa_{t-dt})&=-Z_{t-dt} R^{-1}_{t-dt} E^Q_{t-dt}[dS_t]-Z_{t-dt} E^Q_{t-dt}[dR^{-1}_t dS_t]\notag\\
&=-Z_{t-dt} c^{R^{-1} S}_t dt,
\end{align}
where we have used the martingale property of $S_t$ under the marginal pricing measure $Q$ for the second step. Likewise,
$$f_{\varphi\varphi}(t-dt,x,\varphi_t(x),\kappa_{t-dt}(x))= E_{t-dt}[U''(t,x+\varphi_t(x)dS_t-\kappa_{t-dt}(x)dt)(dS_t)^2],$$
so that evaluation at $X_{t-dt}$ yields
\begin{align}
f_{\varphi\varphi}(t-dt,X_{t-dt},\varphi_t,\kappa_{t-dt})&=-Z_{t-dt} E^Q_{t-dt}[R_{t}^{-1} (dS_t)^2]\notag\\
&=-Z_{t-dt}\left(R^{-1}_{t-dt} E^Q_{t-dt}[(dS_t)^2]-E^Q_{t-dt}[dR_t^{-1}(dS_t)^2]\right)\notag\\
&=-Z_{t-dt} R^{-1}_{t-dt} c^S_t dt.\label{eq:3}
\end{align}
Next,
$$f_{x\kappa}(t-dt,x,\varphi_t(x),\kappa_{t-dt}(x))=E_{t-dt}[U''(t,x+\varphi_t(x)dS_t-\kappa_{t-dt}(x)dt)(-dt)],$$
and in turn
\begin{align}
f_{x\kappa}(t-dt,X_{t-dt},\varphi_t,\kappa_{t-dt})=Z_{t-dt} E^Q_{t-dt}[R^{-1}_{t}dt]&=Z_{t-dt} R^{-1}_{t-dt} dt+Z_{t-dt} E^Q_{t-dt}[dR^{-1}_t dt]\notag\\
&=Z_{t-dt} R^{-1}_{t-dt} dt.\label{eq:4}
\end{align}
Finally: 
$$f_{\kappa\kappa}(t-dt,x,\varphi_t(x),\kappa_{t-dt}(x))=u_1''(t-dt,\kappa_{t-dt})dt+E_{t-dt}[U''(t,x+\varphi_t(x)dS_t-\kappa_{t-dt}(x)dt)(dt)^2].$$
Hence, arguing as above and using the definition of $r_{t-dt}=-u_1'(t-dt,\kappa_{t-dt})/u_1''(t-dt,\kappa_{t-dt})$ as well as \eqref{eq:Zu1}:
\begin{equation}\label{eq:5}
f_{\kappa\kappa}(t-dt,X_{t-dt},\varphi_t,\kappa_{t-dt}) = -Z_{t-dt} r^{-1}_{t-dt} dt - Z_{t-dt} E^Q_{t-dt}[R_t^{-1} (dt)^2]=-Z_{t-dt} r^{-1}_{t-dt} dt. 
\end{equation}
Inserting (\ref{eq:1}--\ref{eq:5}) into \eqref{eq:uxx} then gives
\begin{align*}
U''(t-dt,X_{t-dt})&= -Z_{t-dt} R^{-1}_{t-dt} -Z_{t-dt}\left(b^{R^{-1},Q}_t-R_{t-dt} \frac{(c^{R^{-1} S}_t)^2}{c^S_t}-\frac{r_{t-dt}}{R_{t-dt}^2}\right)dt\\
&= -Z_{t-dt} R^{-1}_{t-dt} -Z_{t-dt}\left(b^{R^{-1},Q}_t-(R_{t}-dR_t) \frac{(c^{R^{-1} S}_t)^2}{c^S_t}-\frac{r_{t}-dr_t}{(R_{t}-dR_t)^2}\right)dt\\
&= -Z_{t-dt} R^{-1}_{t-dt} -Z_{t-dt}\left(b^{R^{-1},Q}_t-R_{t} \frac{(c^{R^{-1} S}_t)^2}{c^S_t}-\frac{r_{t}}{R_{t}^2}\right)dt.
\end{align*}
As $-Z_{t-dt} R^{-1}_{t-dt}=U''(t-dt,X_{t-dt})$ by definition of the indirect risk tolerance and \eqref{eq:z}, it follows that
$$b^{R^{-1},Q}_t=R_t \frac{(c^{R^{-1} S}_t)^2}{c^S_t}+\frac{r_t}{R_t^2}.$$
Now, taking into account that It\^o's formula yields $dR^{-1}_t=-R^{-2}_t dR_t+R_t^{-3}d\langle R \rangle_t$ and in turn $b_t^{R^{-1},Q}=-R_t^{-2}b_t^{R,Q}+R_t^{-3}c_t^R$ as well as $c^{R^{-1} S}_t=-R_t^{-2}c_t^{R S}$, we obtain the following BSDE for the indirect risk tolerance $R_t$:
\begin{equation}\label{eq:BSDE}
b^{R,Q}_t=\frac{1}{R_t}\left(c^R_t-\frac{(c^{R S}_t)^2}{c^S_t}\right)-r_t, \qquad R_T=-\frac{u_2'(X_T)}{u_2''(X_T)},
\end{equation}
where the terminal condition follows directly from the definition of $R_t$. The special cases of utility only from terminal wealth ($u_1(t,x)=0$) or only from intermediate consumption ($u_2(x)=0$) can be dealt with as above, setting either $r_t$, $t<T$, or the terminal value $R_T$ equal to zero.

\begin{myrek}
Equation \eqref{eq:BSDE} is a \emph{quadratic BSDE} for $R_t$. Indeed, suppose the filtration is generated by a $d$-dimensional $Q$-Brownian motion $W_t^Q$ and $dS_t=\sigma_t dW^Q_t$ for an $\mathbb{R}^d$-valued volatility process $\sigma_t$. Then, \eqref{eq:BSDE} takes the form
$$dR_t=\left(\frac{\zeta_t^\top \zeta_t}{R_t}-\frac{(\sigma_t^\top \zeta_t)^2}{R_t \sigma_t^\top \sigma_t}-r_t\right)dt+\zeta_t dW^Q_t, \qquad R_T=-\frac{u_2'(X_T)}{u_2''(X_T)}.$$
\end{myrek}

In the above computations, we have also determined the sensitivity of the optimal portfolio/consumption pair $(\varphi_t,\kappa_t)$ with respect to changes in wealth. Indeed, evaluating \eqref{eq:sensitivity} along the optimal wealth process and accounting for 
(\ref{eq:2}--\ref{eq:5}) as well as $c^{R^{-1} S}_t=-R_t^{-2}c_t^{R S}$, we obtain
\begin{equation}\label{eq:sens}
\varphi_t'=\frac{c^{RS}_t}{R_t c^S_t}, \qquad \kappa'_t=\frac{r_t}{R_t}.
\end{equation}

\subsection{Expected Risk Tolerances and Risk-Tolerance Wealth Processes}\label{ss:B2}

Next, we discuss how the indirect risk tolerance can be represented as the expectation of risk tolerances with respect to future consumption and terminal wealth, computed under a suitable equivalent martingale measure.

\paragraph{Risk-Tolerance Wealth Processes}

First, consider the special case where total risk tolerance 
$$\int_0^T r_t dt+ R_T=\int_0^T -\frac{u_1'(t,\kappa_t)}{u_1''(t,\kappa_t)}dt-\frac{u_2'(X_T)}{u_2''(X_T)}$$ 
can be replicated by dynamic trading in the frictionless market. In the absence of consumption, this means that a \emph{risk-tolerance wealth process} in the sense of Kramkov and S{\^{\i}}rbu \cite{kramkov.sirbu.06} exists. 

If the total risk tolerance can be replicated, $\int_0^T r_t dt + R_T=R_0+\int_0^T\psi_t dS_t$, then our indirect risk tolerance can be represented as the expectation of future risk tolerances with respect to consumption and terminal wealth, computed under under the marginal pricing measure $Q$:
\begin{equation}\label{eq:exprisktolerance}
R_t=E^Q_{t}\left[\int_t^T r_s ds +R_T\right]=E^Q_t\left[\int_t^T -\frac{u_1'(s,\kappa_s)}{u_1''(s,\kappa_s)} ds -\frac{u'_2(X_T)}{u_2''(X_T)}\right].
\end{equation}
Indeed, the $Q$-martingale property of $S_t$ and in turn $\int_0^t \psi_s dS_s$ enables us to rewrite \eqref{eq:exprisktolerance} as $\int_0^t r_s ds +R_t=R_0 +\int_0^t \psi_s dS_s$, so that $R_t$ has $Q$-drift $b_t^{R,Q}=-r_t$. Moreover, this representation also implies $c_t^{R}-(c^{RS}_t)^2/c^S_t=\psi_t^2 c^{S}_t-\psi^2_t c^S_t=0$, thereby showing that the right-hand side of \eqref {eq:exprisktolerance} indeed satisfies the BSDE \eqref{eq:BSDE} for $R_t$. Without intermediate consumption ($r_t=0$), this also shows that the indirect risk tolerance is given by the risk-tolerance wealth process $E^Q[R_T]+\int_0^t \psi_s dS_s$, if the latter exists. The replicability of the total risk tolerance is guaranteed in the important special cases of \emph{complete markets} and \emph{standard utility functions} of exponential or power type:

\begin{enumerate}
\item \emph{Complete Markets}: here, any payoff can be replicated, so in particular this holds for the total risk tolerance. Then, \eqref{eq:exprisktolerance} is computed under the unique equivalent martingale measure for the market at hand. 
\item \emph{Exponential Utilities}: suppose the utilities from terminal wealth and intermediate consumption all have constant absolute risk tolerances: $u_2(x)=-e^{-p_2 x}$ and $u_1(t,x)=-\beta e^{\delta(T-t)} e^{-p_1 x}$ for risk-tolerances $1/p_1,1/p_2>0$, some impatience rate $\delta>0$, and a scaling factor $\beta \geq 0$. Then, the total risk tolerance is constant and therefore evidently replicated by $\psi_t=0$. As a consequence, the indirect risk tolerance is deterministic and given by $R_t=1/p_2+(T-t)1_{\{\beta>0\}}/p_1$. In particular, it coincides with the investors' constant risk-tolerance in the absence of consumption. The sensitivities in \eqref{eq:sens} read as
$\varphi'_t=0$ and $\kappa'_t=1_{\{\beta>0\}}/(T-t+p_1/p_2)$.
\item \emph{Power Utilities}: next, consider utilities with constant \emph{relative} risk tolerance, in the absence of random endowment ($\Psi_t=0$). First, suppose there is no intermediate consumption ($r_t=0$) and utility from terminal wealth has constant relative risk tolerance $1/\gamma>0$, that is, $u_2(x)=x^{1-\gamma}/(1-\gamma)$. In this case, the total risk tolerance is given by the $1/\gamma$-fold of the optimal wealth process, $R_T=X_T/\gamma$, and is therefore evidently replicated by the $1/\gamma$-fold of the optimal trading strategy $\varphi_t$. With intermediate consumption this remains valid, if the utilities $u_1(t,x)$ also have the same constant relative risk tolerance $1/\gamma$, e.g., if they are of the widely used form $u_1(t,x)=\beta e^{\delta(T-t)}x^{1-\gamma}/(1-\gamma)$ for some impatience rate $\delta>0$ and scaling factor $\beta \geq 0$. Then, the risk tolerance with respect to consumption is given by the $1/\gamma$-fold of the latter: $r_t=\kappa_t/\gamma$. As a result, $R_t=X_t/\gamma$ for the wealth process $X_t=x+\int_0^t \varphi_s dS_s -\int_0^t \kappa_s ds$ generated by the optimal portfolio/consumption pair $(\varphi_t,\kappa_t)$. Indeed, $R_t=X_t/\gamma$ has $Q$-drift $-(\kappa_t/\gamma) dt=-r_t dt$, because the risky asset is a $Q$-martingale. As it also satisfies $c^R_t-(c^{RS}_t)^2/c^S_t=0$, it thereby solves the BSDE \eqref{eq:BSDE}. The sensitivities in \eqref{eq:sens} are given by $\varphi'_t=\varphi_t/X_t$ and $\kappa'_t=\kappa_t/X_t$.
\end{enumerate}

\paragraph{Expected Risk-Tolerance Beyond the Replicable Case}

If the total risk tolerance can be replicated by dynamic trading, \eqref{eq:exprisktolerance} shows that the indirect risk tolerance can be interpreted as the expectation of risk tolerances with respect to future consumption and terminal wealth, computed under the investors' marginal pricing measure. Generally, this result is only applicable if the market is complete or the investors' preferences are described by a standard utility function. Nevertheless, the interpretation of the indirect risk tolerance as an expectation of future risk tolerances with respect to consumption and terminal wealth can be transferred to the general case, as we now argue. To this end, decompose the risk tolerance-process $R_t$ as follows:
\begin{equation}\label{eq:gkw}
dR_t=b^{R,Q}_t dt+\frac{c^{RS}_t}{c^S_t}dS_t+dR^\perp_t.
\end{equation}
Then, $R^\perp_t$ is a $Q$-martingale orthogonal to $S_t$, because $d\langle R^\perp, S \rangle_t= c^{RS}_t dt - \frac{c^{RS}_t}{c^S_t} c^S_t dt=0$. Hence, \eqref{eq:gkw} is  -- up to the drift term -- the Galtchouk-Kunita-Watanabe decomposition of the indirect risk tolerance. Now, consider the measure 
$$\widetilde{Q}\sim Q, \quad \mbox{with density process $\widetilde{Z}_t=\E(R^{-1} \mal R^\perp)_t$}.$$
This defines an equivalent martingale measure: as $S_t$ and $\widetilde{Z}_t$ are orthogonal $Q$-martingales, integration by parts shows that $\widetilde{Z}_t S_t$ is a $Q$-martingale and $S_t$ is in turn a $\widetilde{Q}$-martingale. 

In the general case, the equivalent martingale measure $\widetilde{Q}$ replaces the marginal pricing measure $Q$ in determining the indirect risk tolerance as an expectation of future risk tolerances with respect to consumption and terminal wealth:
\begin{equation}\label{eq:general}
R_t=E_t^{\widetilde{Q}}\left[\int_t^T r_s ds +R_T \right].
\end{equation}
To see why \eqref{eq:general} holds true, verify that the right-hand-side of \eqref{eq:general} satisfies the BSDE \eqref{eq:BSDE}. By \eqref{eq:general} and Girsanov' theorem, the drift of $R_t$ under the measure $\widetilde{Q}$ with density process $\widetilde{Z}_t$ is given by
$$-r_t dt= b^{R,\widetilde{Q}}dt=b^{R,Q}_t dt-\langle R, R^{-1} \mal R^\perp \rangle_t=b^{R,Q}_t dt -\frac{1}{R}_t\left(c^R_t-\frac{(c^{RS}_t)^2}{c^S_t}\right)dt.$$
As a consequence, the specification \eqref{eq:general} indeed solves \eqref{eq:BSDE} and therefore provides an interpretation of the risk tolerance-process as the expectation of future risk tolerances under $\widetilde{Q}$.

\section{Martingale Optimality Conditions}\label{sec:martopt}

In this section, we derive conditions that ensure the \emph{approximate} optimality of a family of candidate policies as the spread $\varepsilon_t=\varepsilon \mathcal{E}_t$ becomes small. These sufficient conditions form the basis for the derivations in Appendix~\ref{sec:opt}.
To ease the exposition, we first briefly recall their \emph{exact} counterparts for the frictionless case ($\varepsilon=0$).

\subsection{Frictionless Optimality Conditions}

Without transaction costs ($\varepsilon=0$), a portfolio/consumption pair $(\varphi_t,\kappa_t)$ is optimal if (and essentially only if  \cite{karatzas.shreve.98,karatzas.zitkovic.03}) there exists a process $Z_t$ satisfying the following optimality conditions:
\begin{enumerate}
\item[i)] $Z_t$ is a martingale.
\item[ii)] $Z_t$ is a martingale density, i.e., $Z_t S_t$ is a martingale.
\item[iii)] $Z_t=u_1'(t,\kappa_t)$ for $0 \leq t < T$, and $Z_T=u_2'(X_T(\varphi,\kappa))$.
\end{enumerate}
The first two conditions imply that $Z_t$ is -- up to normalization --  the density of an equivalent martingale measure $Q$. The third links it to the optimal consumption stream and terminal payoff by the usual first-order conditions.

Let us briefly recall why Conditions i)-iii) imply the optimality of $(\varphi_t,\kappa_t)$. For any competing portfolio/consumption pair $(\psi_t,k_t)$, the concavity of the utility functions $u_1(t,\cdot)$ and $u_2(\cdot)$, Condition iii), and Fubini's theorem imply
\begin{align*}
&E\left[u_1(\cdot,k)\mal I_T +u_2(X_T(\psi,k))\right]-E\left[u_1(\cdot,\kappa) \mal I_T+u_2(X_T(\varphi,\kappa))\right]\\
 & \qquad\leq  E\left[u_1'(\cdot,\kappa)(k-\kappa)\mal I_T+u_2'(X_T(\varphi,\kappa) ) (X_T(\psi,k)-X_T(\varphi,\kappa))\right]\\
& \qquad=E\left[Z(k-\kappa)\mal I_T+ Z_T(X_T(\psi,k)-X_T(\varphi,\kappa))\right]\\
& \qquad= Z_0 E^Q\left[(k-\kappa)\mal I_T+(X_T(\psi,k)-X_T(\varphi,\kappa))\right]\\
& \qquad= Z_0 E^Q\left[(\psi-\varphi) \mal S_T\right]=0,
\end{align*}
where the last equality follows from the $Q$-martingale property of $(\psi-\varphi) \mal S_t$, which is a consequence of Conditions i) and ii). This shows that $(\varphi_t,\kappa_t)$ is indeed optimal.

\subsection{Approximate Optimality Conditions with Transaction Costs}

Let us now derive approximate versions of the optimality conditions i)-iii), ensuring the \emph{approximate} optimality of a family $(\varphi_t^\varepsilon,\kappa_t^\varepsilon)_{\varepsilon>0}$ of portfolio consumption pairs, at the leading order $O(\varepsilon^{2/3})$ as the spread $\varepsilon_t=\varepsilon \mathcal{E}_t$ tends to zero.\footnote{Rogers \cite{rogers.04} provides a simple probabilistic argument why this is the relevant order.} As the frictional optimizers converge to their frictionless counterparts $(\varphi_t,\kappa_t)$, it suffices to consider strategies and consumption rates that coincide with $\varphi_t$ resp.\ $\kappa_t$ up to terms of order $o(1)$.

Let $(\varphi_t^\varepsilon,\kappa_t^\varepsilon)_{\varepsilon>0}$ be a family of portfolio/consumption pairs whose optimality we want to verify. As in the frictionless case above, for any family of competitors $(\psi_t^\varepsilon,k_t^\varepsilon)_{\varepsilon>0}$, the concavity of the utilities $u_1(t,\cdot)$ and $u_2(\cdot)$ implies
\begin{equation}\label{eq:concave}
\begin{split}
&E\left[u_1(\cdot,k^\varepsilon) \mal I_T+u_2(X^\varepsilon_T(\psi^\varepsilon,k^\varepsilon))\right]-E\left[u_1(\cdot,\kappa^\varepsilon) \mal I_T+u_2(X^\varepsilon_T(\varphi^\varepsilon,\kappa^\varepsilon))\right]\\
 &\qquad  \leq  E\left[u_1'(\cdot,\kappa^\varepsilon)(k^\varepsilon-\kappa^\varepsilon)\mal I_T+u_2'(X^\varepsilon_T(\varphi^\varepsilon,\kappa^\varepsilon)) (X^\varepsilon_T(\psi^\varepsilon,k^\varepsilon)-X^\varepsilon_T(\varphi^\varepsilon,\kappa^\varepsilon))\right],
\end{split}
\end{equation}
where $X_t^\varepsilon(\psi^\varepsilon,k^\varepsilon)$ and $X_t^\varepsilon(\varphi^\varepsilon,\kappa^\varepsilon)$ are the wealth processes generated by the portfolio/consumption pairs $(\psi_t,k_t)$ resp.\ $(\varphi_t,\kappa_t)$ with transaction costs $\varepsilon_t=\varepsilon \mathcal{E}_t$. Now, suppose we can find frictionless \emph{shadow prices} $S_t^\varepsilon$ evolving in the bid-ask spreads $[S_t-\varepsilon_t,S_t+\varepsilon_t]$, which match the trading prices $S_t \pm\varepsilon_t$ in the original market with transaction costs whenever the respective strategy $\varphi_t^\varepsilon$ trades. Then, the frictional wealth associated to $(\varphi_t^\varepsilon,\kappa_t^\varepsilon)$ evidently coincides with its frictionless counterpart for $S_t^\varepsilon$, i.e., $X^\varepsilon_T(\varphi^\varepsilon,\kappa^\varepsilon)=x+\varphi^\varepsilon \mal S^\varepsilon_T -\kappa^\varepsilon \mal I_T+\Psi_T$. For any other policy $(\psi_t,k_t)$, trading in terms of $S_t^\varepsilon$ rather than with the original bid-ask spread can only increase wealth, because trades are carried out at potentially more favorable prices: $X^\varepsilon_T(\psi^\varepsilon,k^\varepsilon) \leq x+\psi^\varepsilon \mal S^\varepsilon_T -k^\varepsilon \mal I_T+\Psi_T$. Together with \eqref{eq:concave}, this implies:
\begin{equation}\label{eq:shadowest}
\begin{split}
&E\left[u_1(\cdot, k^\varepsilon) \mal I_T +u_2(X^\varepsilon_T(\psi^\varepsilon,k^\varepsilon))\right]-E\left[u_1(\cdot,\kappa^\varepsilon) \mal I_T+u_2(X^\varepsilon_T(\varphi^\varepsilon,\kappa^\varepsilon))\right]\\
 &\qquad  \leq  E\left[u_1'(\cdot,\kappa^\varepsilon)(k^\varepsilon-\kappa^\varepsilon)\mal I_T+u_2'(X^\varepsilon_T(\varphi^\varepsilon,\kappa^\varepsilon)) \left((\psi^\varepsilon-\varphi^\varepsilon) \mal S^\varepsilon_T - (k^\varepsilon-\kappa^\varepsilon)\mal I_T\right)\right].
\end{split}
\end{equation}
Now, suppose we can find a process $Z_t^\varepsilon$ satisfying the following approximate versions of i)-iii) above:
\begin{enumerate}
\item[$\mbox{i}^\varepsilon)$] $Z_t^\varepsilon$ is approximately a martingale, up to a  drift $b^{Z^{\varepsilon}} \mal I_t$ of order $O(\varepsilon^{2/3})$.
\item[$\mbox{ii}^\varepsilon)$] The martingale part $M_t^{Z^\varepsilon}$ of $Z_t^\varepsilon$ is an approximate martingale density, in that the drift rate of $M_t^{Z^\varepsilon} S_t^{\varepsilon}$ is of order $O(\varepsilon^{2/3})$.
\item[$\mbox{iii}^\varepsilon)$] $Z^\varepsilon_t=u'_1(t,\kappa^\varepsilon_t)+O(\varepsilon^{2/3})$ for $0\leq t < T$, and $Z^\varepsilon_T=u_2'(X^\varepsilon_T(\varphi^\varepsilon,\kappa^\varepsilon))+O(\varepsilon^{2/3})$.
\end{enumerate} 
Then, as $k^\varepsilon_t-\kappa^\varepsilon_t=o(1)$ and $\psi^\varepsilon_t-\varphi^\varepsilon_t=o(1)$, Condition $\mbox{iii}^\varepsilon)$ implies that \eqref{eq:shadowest} can be written as
\begin{equation}\label{eq:shadowest2}
\begin{split}
&E\left[u_1(\cdot,k^\varepsilon) \mal I_T +u_2(X^\varepsilon_T(\psi^\varepsilon,k^\varepsilon))\right]-E\left[u_1(\cdot,\kappa^\varepsilon) \mal I_T+u_2(X^\varepsilon_T(\varphi^\varepsilon,\kappa^\varepsilon))\right]\\
 & \qquad \leq  E\left[Z^\varepsilon(k^\varepsilon-\kappa^\varepsilon)\mal I_T+Z^\varepsilon_T \left((\psi^\varepsilon-\varphi^\varepsilon) \mal S^\varepsilon_T -(k^\varepsilon-\kappa^\varepsilon)\mal I_T\right)\right]+o(\varepsilon^{2/3}).
\end{split}
\end{equation}
Let $Q^\varepsilon$ be the measure with density process given by the martingale part $M_t^{Z^\varepsilon}$ of $Z_t^\varepsilon$. Then Fubini's theorem yields
\begin{align*}
E[Z^\varepsilon(k^\varepsilon-\kappa^\varepsilon)\mal I_T]&=Z_0^\varepsilon E^{Q^\varepsilon}[(k^\varepsilon-\kappa^\varepsilon)\mal I_T]+E[((b^{Z^\varepsilon}\mal I)(k^\varepsilon-\kappa^\varepsilon))\mal I_T]\\
&=Z_0^\varepsilon E^{Q^\varepsilon}[(k^\varepsilon-\kappa^\varepsilon)\mal I_T]+o(\varepsilon^{2/3}),
\end{align*}
where we have used $k_t^\varepsilon-\kappa_t^\varepsilon=o(1)$ and $\mbox{i}^\varepsilon)$ for the second step. Likewise, $\mbox{i}^\varepsilon)$ also yields
\begin{align*}
E\left[Z^\varepsilon_T \left((\psi^\varepsilon-\varphi^\varepsilon) \mal S^\varepsilon_T -(k^\varepsilon-\kappa^\varepsilon)\mal I_T\right)\right]&=Z_0^\varepsilon E^{Q^\varepsilon}\left[ \left((\psi^\varepsilon-\varphi^\varepsilon) \mal S^\varepsilon_T -(k^\varepsilon-\kappa^\varepsilon)\mal I_T\right)\right]+o(\varepsilon^{2/3}),
\end{align*}
because $\psi_t^\varepsilon-\varphi_t^\varepsilon$ and $k_t^\varepsilon-\kappa_t^\varepsilon$ are both of order $o(1)$. Combining these two estimates gives
\begin{align*}
&E\left[u_1(\cdot,k^\varepsilon) \mal I_T +u_2(X^\varepsilon_T(\psi^\varepsilon,k^\varepsilon))\right]-E\left[u_1(\cdot,\kappa^\varepsilon) \mal I_T+u_2(X^\varepsilon_T(\varphi^\varepsilon,\kappa^\varepsilon))\right]\\
&\qquad \leq Z_0^\varepsilon E^{Q^\varepsilon}[(\psi^\varepsilon-\varphi^\varepsilon)\mal S^\varepsilon_T]+o(\varepsilon^{2/3}).
\end{align*}
To establish the approximate optimality of the candidate family $(\varphi_t^\varepsilon,\kappa_t^\varepsilon)_{\varepsilon>0}$, it therefore remains to verify that -- given $\mbox{ii}^\varepsilon)$ -- the process $(\psi^\varepsilon-\varphi^\varepsilon) \mal S^\varepsilon_t$ is an approximate $Q^\varepsilon$-martingale, in that the drift rate of $M_t^{Z^\varepsilon}((\psi^\varepsilon-\varphi^\varepsilon) \mal S_t^\varepsilon)$ is of order $o(\varepsilon^{2/3})$. But this readily follows, applying integration by parts twice to obtain
\begin{align*}
M^{Z^\varepsilon}_t ((\psi^\varepsilon-\varphi^\varepsilon) \mal S^\varepsilon_t) &=((\psi^\varepsilon-\varphi^\varepsilon)\mal S^\varepsilon) \mal M^{Z^\varepsilon}_t +M^{Z^\varepsilon}(\psi^\varepsilon-\varphi^\varepsilon) \mal S^\varepsilon_t +(\psi^\varepsilon-\varphi^\varepsilon) \mal \langle M^{Z^\varepsilon},S^\varepsilon \rangle_t \\
&=((\psi^\varepsilon-\varphi^\varepsilon)\mal S^\varepsilon -(\psi^\varepsilon-\varphi^\varepsilon)S^\varepsilon) \mal M^{Z^\varepsilon}_t+(\psi^\varepsilon-\varphi^\varepsilon)\mal (M^{Z^\varepsilon}S^\varepsilon)_t,
\end{align*}
and taking into account that the drift rate of the stochastic integral with respect to the martingale $M_t^{Z^\varepsilon}$ vanishes, whereas the drift rate of the stochastic integral with respect to $M_t^{Z^\varepsilon} S_t^\varepsilon$ is of order $o(\varepsilon^{2/3})$ because of $\mbox{ii}^\varepsilon)$ and because $\psi_t^\varepsilon-\varphi_t^\varepsilon$ is of order $o(1)$.

Summing up, a family of policies $(\varphi_t^\varepsilon,\kappa_t^\varepsilon)$ is indeed approximately optimal if we can find shadow prices $S_t^\varepsilon$ and approximate martingale densities $Z_t^\varepsilon$ satisfying the approximate optimality conditions $\mbox{i}^\varepsilon)$-$\mbox{iii}^\varepsilon)$ above.

\section{Approximate Optimality of the Candidate Policy}\label{sec:opt}

We now verify that the policy proposed in Section \ref{sec:policy} satisfies the approximate optimality conditions $\mbox{i}^\varepsilon)$-$\mbox{iii}^\varepsilon)$ from Appendix~\ref{sec:martopt}, and is therefore indeed optimal at the leading order $O(\varepsilon^{2/3})$ for small transaction costs $\varepsilon_t=\varepsilon \mathcal{E}_t$. In a first step, we construct a candidate shadow price. Subsequently, we put forward a corresponding martingale density, which satisfies $\mbox{i}^\varepsilon)$  and $\mbox{iii}^\varepsilon)$. In a third step, we then show that this martingale density and the candidate shadow price also satisfy $\mbox{ii}^\varepsilon)$.

\textit{Step 1}: define the direct risk tolerance $r_t$ from current consumption and the indirect risk tolerance $R_t$ as in Appendix \ref{sec:risktol}, and let $\varphi_t^\varepsilon$ correspond to the minimal amount of trading necessary to keep the number of risky shares in the random and time-varying no-trade region $[\overline{\mathrm{NT}}_t-\Delta\mathrm{NT}_t,\overline{\mathrm{NT}}_t+\Delta\mathrm{NT}_t]$. By definitions \eqref{eq:deviation} and \eqref{eq:sens}, its midpoint and halfwidth are given by
\begin{gather*}
\overline{\mathrm{NT}}_t=\varphi_t+\varphi'_t (X_t^\varepsilon(\varphi^\varepsilon,\kappa^\varepsilon)-X_t(\varphi,\kappa))=\varphi_t +\frac{c^{RS}_t}{R_t c^S_t} (X_t^\varepsilon(\varphi^\varepsilon,\kappa^\varepsilon)-X_t(\varphi,\kappa)),\\
\Delta\mathrm{NT}_t=\left(\frac{3R_t}{2} \frac{d\langle \varphi \rangle_t}{d\langle S \rangle_t} \varepsilon_t\right)^{1/3}=\left(\frac{3R_t}{2} \frac{c^{\varphi}_t}{c^S_t} \varepsilon_t\right)^{1/3},
\end{gather*}
respectively. This no-trade region is not symmetric around the frictionless optimizer $\varphi_t$. To nevertheless construct a shadow price similarly as in the symmetric exponential case \cite{kallsen.muhlekarbe.12}, decompose the deviation $\Delta\varphi_t=\varphi_t^\varepsilon-\varphi_t$ of the frictional position $\varphi_t^\varepsilon$ from its frictionless counterpart into two parts: $\Delta\varphi_t=\overline{\Delta\varphi}_t+\widetilde{\Delta\varphi}_t$. Here, the first term $\overline{\Delta\varphi}_t=\overline{\mathrm{NT}}_t-\varphi_t$ measures the deviation of the midpoint $\overline{\mathrm{NT}}_t$ of the no-trade region from the frictionless target position $\varphi_t$. The second term $\widetilde{\Delta\varphi}_t=\varphi_t^\varepsilon-\overline{\mathrm{NT}}_t$ in turn describes the deviation of the frictional position $\varphi_t^\varepsilon$ from the midpoint $\overline{\mathrm{NT}}_t$. Note that this is an It\^o process, reflected to remain in the random and time-varying but symmetric interval $[-\Delta\mathrm{NT}_t,\Delta\mathrm{NT}_t]$. With this notation, define\footnote{Note that these definitions match the ones for the case of exponential utility \cite{kallsen.muhlekarbe.12}, where the indirect risk tolerance $R_t$ is constant, and the frictionless optimizer coincides with the midpoint of the no-trade region.}
$$S^\varepsilon_t =S_t+\Delta S_t = S_t+ \alpha_t \widetilde{\Delta\varphi}_t^3 -\gamma_t \widetilde{\Delta\varphi}_t,$$
for 
$$\alpha_t=\frac{1}{3R_t}\frac{c_t^S}{c_t^\varphi}, \qquad \gamma_t=\left(\frac{9}{4R_t}\frac{c_t^S}{c_t^\varphi}\varepsilon_t^2\right)^{1/3}. $$
One readily verifies that the process $S_t^\varepsilon$ takes values in the bid-ask spread $[S_t-\varepsilon_t,S_t+\varepsilon_t]$. Moreover, as $\alpha_t(\pm\Delta\mathrm{NT}_t)^3- \gamma_t(\pm\Delta\mathrm{NT}_t)=\mp\varepsilon_t$, it coincides with the bid resp.\ ask price whenever the policy $\varphi_t^\varepsilon$ prescribes the purchase resp.\ sale of risky shares after reaching the boundaries of the no-trade region for $\widetilde{\Delta\varphi}_t=\pm\Delta\mathrm{NT}_t$. Consequently, $S_t^\varepsilon$ is a valid candidate shadow price process. 

Next, note that $3\alpha_t(\Delta\mathrm{NT}_t)^2-\gamma_t=0$. As a result, integration by parts and It\^o's formula give
\begin{align}
\Delta S_t-\Delta S_0 &=\alpha \mal \widetilde{\Delta\varphi}_t^3 + \widetilde{\Delta\varphi}^3 \mal \alpha_t +\langle \alpha, \widetilde{\Delta\varphi}^3 \rangle_t -\gamma \mal \widetilde{\Delta\varphi}_t -\widetilde{\Delta\varphi} \mal \gamma_t - \langle \gamma, \widetilde{\Delta\varphi} \rangle_t \notag \\
&= (3\alpha \widetilde{\Delta\varphi}^2-\gamma) \mal \widetilde{\Delta\varphi}_t + (3\alpha \widetilde{\Delta\varphi}) \mal \langle \widetilde{\Delta\varphi} \rangle_t \notag  \\
& \quad + \widetilde{\Delta\varphi}^3 \mal \alpha_t - \widetilde{\Delta\varphi} \mal \gamma_t +(3\widetilde{\Delta\varphi}^2)\mal \langle \alpha, \widetilde{\Delta\varphi} \rangle_t -\langle \gamma , \widetilde{\Delta\varphi} \rangle_t \notag \\
&= -(3\alpha \widetilde{\Delta\varphi}^2-\gamma) \mal (\varphi+\overline{\Delta\varphi})_t + (3\alpha \widetilde{\Delta\varphi}) \mal \langle \varphi+\overline{\Delta\varphi} \rangle_t \notag \\
& \quad + \widetilde{\Delta\varphi}^3 \mal \alpha_t - \widetilde{\Delta\varphi} \mal \gamma_t -(3\widetilde{\Delta\varphi}^2)\mal \langle \alpha, \varphi+\overline{\Delta\varphi} \rangle_t +\langle \gamma , \varphi+\overline{\Delta\varphi} \rangle_t. \label{eq:ito}
\end{align}
Here, we have used for the last equality that $\varphi_t^\varepsilon=\varphi_t+\overline{\Delta\varphi}_t+\widetilde{\Delta\varphi}_t$ only moves on the set $\widetilde{\Delta\varphi}_t=\pm\Delta\mathrm{NT}_t$ where $3\alpha_t\widetilde{\Delta\varphi}_t^2-\gamma_t=0$, and that $\widetilde{\Delta\varphi}_t=\varphi_t^\varepsilon-\varphi_t-\overline{\Delta\varphi}_t$ only differs from $-(\varphi_t+\overline{\Delta\varphi}_t)$ by a finite variation term. Given that the risky asset $S_t$, the frictionless optimizer $(\varphi_t,\kappa_t)$, the indirect risk tolerance $R_t$, the transaction cost process $\varepsilon_t=\varepsilon \mathcal{E}_t$, and their local quadratic (co-)variations all follow sufficiently regular It\^o processes, the above representation shows that this property is passed on to the coefficient processes $\alpha_t$ and $\gamma_t$ as well as to $\Delta S_t$.  Moreover, \eqref{eq:ito} as well as $\widetilde{\Delta\varphi}_t=O(\varepsilon^{1/3})$, $\gamma_t=O(\varepsilon^{2/3})$, $\sqrt{c^\gamma_t}=O(\varepsilon^{2/3})$, $\alpha_t=O(1)$, and $\sqrt{c^\alpha_t}=O(1)$ yield that the diffusion coefficient of $\Delta S_t$ is given by 
\begin{equation}\label{eq:c0}
\sqrt{c^{\Delta S}_t}=O(\varepsilon^{2/3})+O(\varepsilon^{2/3})\sqrt{c_t^{\overline{\Delta\varphi}}},
\end{equation}
and integration by parts shows that its counterpart for 
\begin{equation}\label{eq:differ}
\overline{\Delta\varphi}_t=\frac{c^{RS}_t}{R_t c^S_t}(X^\varepsilon_t(\varphi^\varepsilon,\kappa^\varepsilon)-X_t(\varphi,\kappa))=\frac{c^{RS}_t}{R_t c^S_t}\left( \Delta \varphi \mal S_t+(\varphi+\Delta\varphi) \mal \Delta S_t -(\kappa^\varepsilon-\kappa) \mal I_t\right)
\end{equation}
in turn satisfies
\begin{align}
\sqrt{c_t^{\overline{\Delta\varphi}}} &=O(1)(X^\varepsilon_t(\varphi^\varepsilon,\kappa^\varepsilon)-X_t(\varphi,\kappa))+O(1)\Delta\varphi_t+O(1)(\varphi_t+\Delta\varphi_t)\sqrt{c^{\Delta S}_t}\notag\\
&= O(1)(X^\varepsilon_t(\varphi^\varepsilon,\kappa^\varepsilon)-X_t(\varphi,\kappa))+O(\varepsilon^{1/3})\notag\\
&\qquad \qquad +O(\varepsilon^{2/3})\sqrt{c^{\overline{\Delta\varphi}}_t}+O(\varepsilon^{2/3})(X^\varepsilon_t(\varphi^\varepsilon,\kappa^\varepsilon)-X_t(\varphi,\kappa))\sqrt{c^{\overline{\Delta\varphi}}_t}\label{eq:cvarphi}\\
&=O(1)+O(\varepsilon^{2/3})\sqrt{c^{\overline{\Delta\varphi}}_t},\notag
\end{align}
where we have used 
\begin{equation}\label{eq:dv}
\Delta\varphi_t=\overline{\Delta\varphi}_t+\widetilde{\Delta\varphi}_t=O(1)(X^\varepsilon_t(\varphi^\varepsilon,\kappa^\varepsilon)-X_t(\varphi,\kappa))+O(\varepsilon^{1/3})
\end{equation}
for the second step.\footnote{For the third step, we assume without loss of generality that the difference $X^\varepsilon_t(\varphi^\varepsilon,\kappa^\varepsilon)-X_t(\varphi,\kappa)$ between the frictional and frictionless wealth processes is of order $O(1)$ for small transaction costs $\varepsilon$. If this does not hold, one can instead consider a modified policy that is stopped if $X^\varepsilon_t(\varphi^\varepsilon,\kappa^\varepsilon)-X_t(\varphi,\kappa)$ exceeds some fixed threshold; a fortiori, it then turns out that the threshold is not hit for sufficiently small transaction costs, cf.~\eqref{eq:c7} below.} Hence,
\begin{equation}\label{eq:c1}
\sqrt{c^{\overline{\Delta\varphi}}_t} = O(1),
\end{equation}
so that, by \eqref{eq:c0}:
\begin{equation}\label{eq:c2}
\sqrt{c^{\Delta S}_t}=O(\varepsilon^{2/3}).
\end{equation}
Together with \eqref{eq:ito}, the estimate in \eqref{eq:c1} yields 
\begin{equation}\label{eq:c5}
b^{\Delta S}_t =O(\varepsilon^{1/3})+O(\varepsilon^{2/3})b^{\overline{\Delta\varphi}}_t.
\end{equation}
Integrating \eqref{eq:differ} by parts and inserting the definition of $\kappa^\varepsilon_t$ from \eqref{eq:consumption}, it therefore follows that
\begin{align}
b^{\overline{\Delta\varphi}}_t &= O(1)(X^\varepsilon_t(\varphi^\varepsilon,\kappa^\varepsilon)-X_t(\varphi,\kappa))+O(1)\Delta\varphi_t+O(1)(\varphi_t+\Delta\varphi_t)b^{\Delta S}_t\notag\\
&= O(1)(X^\varepsilon_t(\varphi^\varepsilon,\kappa^\varepsilon)-X_t(\varphi,\kappa))+O(\varepsilon^{1/3})+O(\varepsilon^{2/3})(X^\varepsilon_t(\varphi^\varepsilon,\kappa^\varepsilon)-X_t(\varphi,\kappa))b^{\overline{\Delta\varphi}}_t +O(\varepsilon^{2/3})b^{\overline{\Delta\varphi}}_t\label{eq:bvarphi}\\
&= O(1)+O(\varepsilon^{2/3})b^{\overline{\Delta\varphi}}_t,\notag
\end{align}
where we have used \eqref{eq:dv} and \eqref{eq:c5} for the second equality.\footnote{For the third step, we again assume without loss of generality that $X^\varepsilon_t(\varphi^\varepsilon,\kappa^\varepsilon)-X_t(\varphi,\kappa)$ is of order $O(1)$.}
As a consequence:
\begin{equation}\label{eq:c3}
b^{\overline{\Delta\varphi}}_t =O(1),
\end{equation}
and therefore, in view of \eqref{eq:c5}:
\begin{equation}\label{eq:c4}
b^{\Delta S}_t =O(\varepsilon^{1/3}).
\end{equation}
Now notice that \eqref{eq:c2} and \eqref{eq:c4} imply
\begin{align*}
X^\varepsilon_t(\varphi^\varepsilon,\kappa^\varepsilon)-X_t(\varphi,\kappa)&= O(\varepsilon^{1/3})+O(1)(X^\varepsilon(\varphi^\varepsilon,\kappa^\varepsilon)-X(\varphi,\kappa))\mal S_t\\
&\qquad +O(1)(X^\varepsilon(\varphi^\varepsilon,\kappa^\varepsilon)-X(\varphi,\kappa))\mal \Delta S_t+O(1)(X^\varepsilon(\varphi^\varepsilon,\kappa^\varepsilon)-X(\varphi,\kappa))\mal I_t.
\end{align*}
Again using \eqref{eq:c2} and \eqref{eq:c4}, Gronwall's lemma therefore shows\footnote{In particular, this difference remains below a given threshold for sufficiently small $\varepsilon$, so that a potential stopping barrier is never hit.} 
\begin{equation}\label{eq:c7}
X^\varepsilon_t(\varphi^\varepsilon,\kappa^\varepsilon)-X_t(\varphi,\kappa)= O(\varepsilon^{1/3}),
\end{equation}
 and in turn $\overline{\Delta\varphi}_t=O(\varepsilon^{1/3})$. Moreover, $b^{\overline{\Delta\varphi}}_t=O(\varepsilon^{1/3})$ as well as $\sqrt{c^{\overline{\Delta\varphi}}_t}=O(\varepsilon^{1/3})$ by \eqref{eq:bvarphi} as well as \eqref{eq:cvarphi}, respectively. Together with \eqref{eq:ito}, it follows that the drift rate $b^{\Delta S}_t$ of $\Delta S_t$ satisfies
\begin{equation}\label{eq:drift}
b_t^{\Delta S}=3\alpha_t \widetilde{\Delta\varphi}_t c_t^\varphi + O(\varepsilon^{2/3}).
\end{equation}

\par\medskip

\textit{Step 2}: Set $\Delta\kappa_t=\kappa_t^\varepsilon-\kappa_t$ and define $\Delta X_t= \Delta\varphi \mal S_t -\Delta\kappa \mal I_t$. At the leading order $O(\varepsilon^{1/3})$, this process measures the difference between the frictionless optimal wealth process $X_t(\varphi,\kappa)=x+ \varphi \mal S_t-\kappa \mal I_t+\Psi_t$ and its frictional counterpart $X_t^\varepsilon(\varphi^\varepsilon,\kappa^\varepsilon)=x+\varphi^\varepsilon \mal S_t^\varepsilon-\kappa^\varepsilon \mal I_t+\Psi_t$. Indeed, 
$$x+\varphi^\varepsilon\mal S_t^\varepsilon-\kappa^\varepsilon \mal I_t -(x+\varphi \mal S_t-\kappa \mal I_t)= \Delta \varphi \mal S_t+\varphi^\varepsilon \mal \Delta S_t -\Delta\kappa \mal I_t= \Delta\varphi \mal S_t -\Delta\kappa \mal I_t+O(\varepsilon^{2/3}),$$
because integration by parts gives 
$$\varphi^\varepsilon \mal \Delta S_t=\Delta\varphi \mal \Delta S_t+\varphi_t \Delta S_t -\varphi_0 \Delta S_0 -\Delta S \mal \varphi_t - \langle \varphi, \Delta S \rangle_t=O(\varepsilon^{2/3})$$
as $\Delta\varphi=O(\varepsilon^{1/3})$ and the drift and diffusion coefficients of $\Delta S_t$ are of order $O(\varepsilon^{1/3})$ and $O(\varepsilon^{2/3})$, respectively, by Step 1. In particular, it follows that 
\begin{equation}\label{eq:diffX}
X^\varepsilon_t(\varphi^\varepsilon,\kappa^\varepsilon)-X_t(\varphi,\kappa)=\Delta X_t+O(\varepsilon^{2/3}).
\end{equation}

Now, define $Z^\varepsilon_t=(1-\Delta X_t/R_t)Z_t$, where $Z_t$ denotes the density process of the frictionless marginal pricing measure $Q$ up to normalization.\footnote{For exponential utilities with constant indirect risk tolerance $R_t$, this definition reduces to the one in \cite{kallsen.muhlekarbe.12}.}  Then, $Z_t^\varepsilon$ satisfies the approximate optimality conditions $\mbox{i}^\varepsilon)$ and $\mbox{iii}^\varepsilon)$. Indeed, Taylor expansion, the frictionless optimality condition iii), as well as
$$\Delta\kappa_t=\kappa_t^\varepsilon-\kappa_t=\frac{r_t}{R_t}\Delta X_t +O(\varepsilon^{2/3}) \quad \mbox{and} \quad  r_t=-\frac{u'_1(t,\kappa_ t)}{u_1''(t,\kappa_t)},$$
(where the first representation follows from \eqref{eq:diffX} and \eqref{eq:sens}) give
\begin{align*}
u'_1(t,\kappa^\varepsilon_t) = u'_1(t,\kappa_t)+\Delta\kappa_t u''_1(t,\kappa_t)+O(\varepsilon^{2/3})&=Z_t(1+\Delta \kappa_t u''_1(t,\kappa_t)/u'_1(t,\kappa_t))+O(\varepsilon^{2/3})\\
&=Z^\varepsilon_t+O(\varepsilon^{2/3}).
\end{align*}
Likewise and also taking into account the terminal condition for $R_T$, one obtains
\begin{align*}
u'_2(X^\varepsilon_T(\varphi^\varepsilon,\kappa^\varepsilon))=u'_2(X_T(\varphi,\kappa))+\Delta X_T u''_2(X_T(\varphi,\kappa))+O(\varepsilon^{2/3})&=Z_T(1-\Delta X_T/R_T)+O(\varepsilon^{2/3})\\
&=Z^\varepsilon_T+O(\varepsilon^{2/3}).
\end{align*}
Hence, the process $Z_t^\varepsilon$ satisfies $\mbox{iii}^{\varepsilon})$. Let us now check that it also verifies $\mbox{i}^\varepsilon)$. To this end, first recall that $Z_t$ is a martingale by the frictionless optimality condition i). Hence, integration by parts shows
\begin{equation}\label{eq:mart}
Z_t^\varepsilon \cong -Z \mal (\Delta X_t/R_t) -\langle Z, \Delta X/R \rangle_t,
\end{equation}
where $\cong$ denotes equality up to a process with of order $O(\varepsilon^{2/3})$ (here, the remainder in fact has zero drift). By It\^o's formula, and as $\Delta X_t=\Delta\varphi\mal S_t-\Delta\kappa\mal I_t$ and $\Delta\kappa_t=\frac{r_t}{R_t}\Delta X_t+O(\varepsilon^{2/3})$:
\begin{align}
\frac{\Delta X_t}{R_t}&=-\frac{\Delta X}{R^2} \mal R_t +\frac{\Delta X}{R^3} \mal \langle R \rangle_t +\frac{\Delta\varphi}{R} \mal S_t -\frac{\Delta \kappa}{R} \mal I_t - \frac{\Delta\varphi}{R^2} \mal \langle R,S \rangle_t \notag\\
&\cong \frac{\Delta X}{R^2} \mal \left(-R_t+\frac{1}{R} \mal \langle R \rangle_t -r\mal I_t\right)+\frac{\Delta\varphi}{R}\mal \left(S_t-\frac{1}{R} \mal \langle R,S \rangle_t\right),\label{eq:xdr}
\end{align}
where $\cong$ again refers to equality up to a process with drift of order $O(\varepsilon^{2/3})$. Inserting this into \eqref{eq:mart} and writing the density process of the frictionless marginal pricing measure $Q$ as $Z_t/Z_0=\E(N)_t$ leads to
\begin{align*}
Z_t^\varepsilon &\cong \frac{Z \Delta X}{R^2} \mal \left(R_t+\langle N,R\rangle_t -\frac{1}{R} \mal \langle R \rangle_t  +r \mal I_t \right)- \frac{Z\Delta\varphi}{R} \mal \left(S_t+\langle N,S \rangle_t- \frac{1}{R}\mal \langle R,S \rangle_t \right) \\
&\cong \frac{Z\Delta X}{R^2}\left(b^{R,Q}-\frac{c^R}{R}+r\right) \mal I_t-\frac{Z \Delta\varphi}{R}\left(b^{S,Q}-\frac{c^{RS}}{R}\right)\mal I_t.
\end{align*}
Now, recall that the deviation $\Delta\varphi_t=\overline{\Delta\varphi}_t+\widetilde{\Delta\varphi}_t$ of the frictional position from its frictionless counterpart is composed of two parts. The first, which represents the shift of the midpoint of the no-trade region due to the past effects of transaction costs, is an It\^o process with drift and diffusion coefficients of order $O(\varepsilon^{1/3})$. In contrast, the second term measuring deviations from the midpoint of the no-trade region, also involves reflection off the boundaries $\pm\Delta\mathrm{NT}_t=O(\varepsilon^{1/3})$. Hence, this process is not only small, but also oscillates quickly around its mean zero. Therefore, it can be neglected at the leading order in the time average above. As a result, replacing $\Delta\varphi_t$ by $\overline{\Delta\varphi}_t=\frac{c^{RS}_t}{R_t c^S_t}\Delta X_t+O(\varepsilon^{2/3})$ (cf.\ \eqref{eq:diffX}):
$$Z_t^\varepsilon \cong  \frac{Z\Delta X}{R^2}\left(b^{R,Q}-\frac{1}{R}\left(c^R-\frac{(c^{RS})^2}{c^S}\right)+r -\frac{c^{RS}}{c^S} b^{S,Q}\right)\mal I_t.$$
The frictionless price process $S_t$ is a martingale under the marginal pricing measure $Q$ with density process $Z_t/Z_0$ by the frictionless optimality condition ii); hence its $Q$-drift rate vanishes, $b_t^{S,Q}=0$. Together with the BSDE \eqref{eq:BSDE} for the indirect risk tolerance $R_t$, this shows that that drift of $Z_t^\varepsilon$ is indeed of order $O(\varepsilon^{2/3})$ as required by $\mbox{i}^\varepsilon)$.

\par\medskip

\textit{Step 3}: to establish the leading-order optimality of the proposed policy $(\varphi_t^\varepsilon,\kappa_t^\varepsilon)$, it remains to verify the approximate optimality condition $\mbox{ii}^\varepsilon)$, i.e., that $M_t^{Z^\varepsilon} S^\varepsilon$ is approximately a martingale for the martingale part $M_t^{Z^\varepsilon}$ of $Z_t^\varepsilon$. To see this, first notice that integration by parts and the martingale property of $M_t^{Z^\varepsilon}$ yield
\begin{align*} 
M_t^{Z^\varepsilon} S_t^\varepsilon  &\cong M^{Z^\varepsilon} \mal S_t^\varepsilon + \langle M^{Z^\varepsilon}, S^\varepsilon \rangle_t 
\cong Z^\varepsilon \mal S_t^\varepsilon +\langle Z^\varepsilon , S^\varepsilon \rangle_t\\
&= Z(1-\Delta X/R) \mal (S_t+\Delta S_t) +\langle Z(1-\Delta X/R), S+\Delta S \rangle_t \\
&\cong Z(1-\Delta X/R) \mal S_t + Z \mal \Delta S_t + \langle Z(1-\Delta X/R), S \rangle_t\\
& =Z(1-\Delta X/R) \mal (S_t+\langle N, S \rangle_t) + Z \mal \Delta S_t -Z \mal \langle \Delta X/R, S \rangle_t,
\end{align*}
where $\cong$ once more denotes equality up to a process with drift of order $O(\varepsilon^{2/3})$. Here, the second step uses that $Z_t^\varepsilon$ only differs from its martingale part $M_t^{Z^\varepsilon}$ by a finite variation drift of order $O(\varepsilon^{2/3})$ as verified above. For the fourth step, we have used that the drift and diffusion coefficients of $\Delta S_t$ are of order $O(\varepsilon^{1/3})$ resp.\ $O(\varepsilon^{2/3})$, and that $\Delta X_t$ is of order $O(\varepsilon^{1/3})$. Now, inserting representation \eqref{eq:xdr} for $\Delta X_t/R_t$ and using that $S_t+\langle N,S \rangle_t$ is a martingale by the frictionless optimality condition ii) and Girsanov's theorem, it follows that
\begin{align*} 
M_t^{Z^\varepsilon} S_t^\varepsilon &\cong Z \mal \left(\Delta S_t-\frac{\Delta X}{R^2} \mal \left(\langle R,S\rangle_t-\frac{c^{RS}}{c^S} \mal \langle S \rangle_t \right)-\frac{\Delta\varphi}{R} \mal \langle S \rangle_t\right)\\
&\cong Z \mal \left(b^{\Delta S}-\frac{c^{S}}{R} \widetilde{\Delta\varphi}\right)\mal I_t \cong 0,
\end{align*}
where we have again replaced $\Delta\varphi_t$ in the time average by $\overline{\Delta\varphi}_t=\frac{c^{RS}_t}{R_t c^S_t}\Delta X_t+O(\varepsilon^{2/3})$ in the second step, whereas the last step is a consequence of \eqref{eq:drift}. In summary, the approximate optimality conditions $\mbox{i}^\varepsilon)$-$\mbox{iii}^\varepsilon)$ are satisfied, so that the policy $(\varphi_t^\varepsilon,\kappa_t^\varepsilon)$ from Section \ref{sec:policy} is indeed approximately optimal.

\section{Computation of the Utility Loss}\label{sec:uloss}

We now turn to the welfare effects of small transaction costs reported in Section~\ref{sec:welfare}. By definition of $\Delta\kappa_t=\kappa_t^\varepsilon-\kappa_t$, the consumption adjustment due to small transaction costs can be written as $\Delta\kappa_t=\frac{r_t}{R_t}\Delta X_t+O(\varepsilon^{2/3})$, where $\Delta X_t=\Delta \varphi \mal S_t -\Delta \kappa \mal I_t$ as above. Then, Taylor expanding $u_1(t,\cdot)$, using Fubini's theorem, and inserting the frictionless optimality condition iii) as well as the definition of $r_t$, it follows that
\begin{align*}
E[u_1(\cdot,\kappa^\varepsilon)\mal I_T]-E\left[u_1(\cdot,\kappa) \mal I_T \right]&=\left(E[u_1'(\cdot,\kappa)\Delta\kappa]+\frac{1}{2}E[u_1''(\cdot,\kappa)\Delta\kappa^2]\right)\mal I_T+O(\varepsilon)\\
&=Z_0 \left(E^Q[\Delta\kappa \mal I_T]-\frac{1}{2} E^Q\left[r\frac{\Delta X^2}{R^2} \mal I_T\right]\right)+O(\varepsilon),
\end{align*}
for the frictionless marginal pricing measure $Q$ with density process given by $Z_t/Z_0$. Similarly, for the utility from terminal wealth, expand $u_2(\cdot)$ and use the frictionless optimality condition iii) as well as $X_t^\varepsilon(\varphi^\varepsilon,\kappa^\varepsilon)-X_t(\varphi,\kappa)=\Delta X_t+O(\varepsilon^{2/3})$ and the terminal condition for $R_T$:
\begin{align*}
E[u_2(X^\varepsilon_T(\varphi^\varepsilon,\kappa^\varepsilon))]-E[u_2(X_T(\varphi,\kappa))] &= E[u_2'(X_T(\varphi,\kappa))(X^\varepsilon_T(\varphi^\varepsilon,\kappa^\varepsilon)-X_T(\varphi,\kappa))]\\
&\qquad +\frac{1}{2}E[u_2''(X_T(\varphi,\kappa)(X^\varepsilon_T(\varphi^\varepsilon,\kappa^\varepsilon)-X_T(\varphi,\kappa))^2]+O(\varepsilon)\\
&= Z_0 E^Q[\Delta\varphi \mal S_T+(\varphi+\Delta\varphi) \mal \Delta S_T-\Delta\kappa \mal I_T]\\
&\qquad -\frac{1}{2} Z_0 E^Q\left[\frac{\Delta X^2_T}{R_T}\right]+O(\varepsilon).
\end{align*}
As the risky asset $S_t$ is a $Q$-martingale by the frictionless optimality condition ii), the $Q$-expectation of $\Delta\varphi \mal S_T$ vanishes. Moreover, by, e.g., \cite[Theorem 3.10.(v)]{karatzas.zitkovic.03} the initial value $Z_0$ of the dual martingale density coincides with the derivative $U'(x)$ of the frictionless value function with respect to wealth (see also \eqref{eq:hyp}). Combining this with the two estimates above, the difference between the utilities obtained by applying the approximately optimal policy $(\varphi_t^\varepsilon,\kappa_t^\varepsilon)$ with transaction costs and the frictionless optimizer $(\varphi_t,\kappa_t)$ is therefore found to be 
\begin{align*}
U^\varepsilon(x)-U(x)=U'(x)\left(-\frac{1}{2} E^Q\left[r\frac{\Delta X^2}{R^2} \mal I_T +\frac{\Delta X^2_T}{R_T}\right]+E^Q[(\varphi+\Delta\varphi) \mal \Delta S_T] \right)+O(\varepsilon).
\end{align*}
Here, the first term also arises when computing the utility derived from $(\varphi_t^\varepsilon,\kappa_t^\varepsilon)$ traded at the mid price $S_t$ rather than the shadow price $S_t^\varepsilon=S_t+\Delta S_t$. Consequently, it measures the displacement loss incurred by deviating from the frictionless optimizer $(\varphi_t,\kappa_t)$. The second term represents the additional losses directly induced by the trading costs. Let us first focus on the displacement loss. Integration by parts and the definition of $\Delta X_t$ yield
$$\Delta X_t^2 = 2\Delta X \Delta \varphi \mal S_t - 2 \Delta X \Delta \kappa \mal I_t + \Delta \varphi^2 \mal \langle S \rangle_t.$$
Moreover, It\^o's formula gives $R^{-1}_t-R^{-1}_0=-R^{-2} \mal R_t +R^{-3} \mal \langle R \rangle_t$. Integrating by parts again, inserting $\Delta\kappa_t=\frac{r_t}{R_t}\Delta X_t+O(\varepsilon^{2/3})$, and using that $S_t$ is a $Q$-martingale it follows that
\begin{align*}
E^Q\left[\frac{\Delta X^2_T}{R_T}\right] &= E^Q\left[\Delta X^2 \mal R^{-1}_T +R^{-1} \mal \Delta X^2_T+ \langle R^{-1}, \Delta X^2 \rangle_T \right]\\
&=E^Q\Big[-\frac{\Delta X^2}{R^2} \mal R_T +\frac{\Delta X^2}{R^3} \mal \langle R,R \rangle_T -\frac{2\Delta X^2 r}{R^2} \mal I_T+\frac{\Delta\varphi^2}{R} \mal \langle S\rangle_T\\
&\qquad\qquad  -\frac{2\Delta X \Delta\varphi}{R^2} \mal \langle R,S \rangle_T\Big] +O(\varepsilon).
\end{align*}
The argument of this expectation has $Q$-drift
\begin{align*}
\frac{\Delta X^2}{R^2} \left(-b^{R,Q}+\frac{1}{R}c^R-2r\right)\mal I_T+\left(\frac{\Delta\varphi^2}{R} c^S-\frac{2\Delta X \Delta \varphi}{R^2} c^{RS}\right)\mal I_T.
\end{align*}
To proceed, extend the ``averaging'' argument from Step 2 of Section \ref{sec:opt} by noticing that -- at the leading order -- all occurrences of the oscillatory part $\widetilde{\Delta\varphi}_t$ of $\Delta\varphi_t=\overline{\Delta\varphi}_t+\widetilde{\Delta\varphi}_t$ can be replaced in the above time integrals by their expectations under the uniform distribution on $[-\Delta\mathrm{NT}_t,\Delta\mathrm{NT}_t]$ (compare \cite{rogers.04,goodman.ostrov.10}; also see Section \ref{sec:tv} for more details). More specifically, approximating $\widetilde{\Delta\varphi}_t$ by $0$ and $\widetilde{\Delta\varphi}_t^2$ by $\Delta\mathrm{NT}_t^2/3$, the above $Q$-drift can be rewritten as  
\begin{align*}
&\frac{\Delta X^2}{R^2} \left(-b^{R,Q}+\frac{1}{R}c^R-2r\right)\mal I_T+\left(\frac{\overline{\Delta\varphi}^2}{R} c^S+\frac{(\Delta\mathrm{NT})^2}{3R} c^S-\frac{2\Delta X \overline{\Delta\varphi}}{R^2} c^{RS}\right)\mal I_T+O(\varepsilon)\\
&\qquad =\frac{\Delta X^2}{R^2} \left(-b^{R,Q}+\frac{1}{R}\left(c^R-\frac{(c^{RS})^2}{c^S}\right)-2r\right)\mal I_T+\frac{(\Delta\mathrm{NT})^2}{3R} c^S \mal I_T+O(\varepsilon),
\end{align*}
where we have used $\overline{\Delta\varphi}_t=\frac{c_t^{RS}}{R_t c_t^S}\Delta X_t+O(\varepsilon^{2/3})$ to obtain the second equality. Taking into account the BSDE \eqref{eq:BSDE} for $R_t$, it therefore follows that
$$E^Q\left[\frac{\Delta X^2_T}{R_T}\right] = E^Q\left[-r\frac{\Delta X^2}{R^2}\mal I_T+\frac{(\Delta\mathrm{NT})^2}{3R} \mal \langle S \rangle_T\right]+O(\varepsilon).$$
The displacement loss is in turn given by
$$-\frac{1}{2}E^Q\left[r\frac{\Delta X^2}{R^2}\mal I_T+\frac{\Delta X^2_T}{R_T}\right] = -\frac{1}{6} E^Q\left[\frac{(\Delta\mathrm{NT})^2}{R} \mal \langle S \rangle_T\right]+O(\varepsilon).$$

Next, consider the transaction cost loss $E^Q[\varphi \mal \Delta S_T]+E^Q[\Delta\varphi \mal \Delta S_T]$. As $\Delta S_t=O(\varepsilon)$, we have
$$\varphi \mal \Delta S_t = \varphi_t \Delta S_t-\varphi_0\Delta S_0 -\Delta S \mal \varphi_t -\langle \Delta S, \varphi \rangle_t =-\langle \Delta S, \varphi \rangle_t +O(\varepsilon).$$
Representation \eqref{eq:ito} for $\Delta S_t$ shows that $-\langle \Delta S,\varphi \rangle_t=(3\alpha \widetilde{\Delta\varphi}^2 -\gamma)\mal \langle \varphi \rangle_t+O(\varepsilon)$, so that it follows from the definition of $\alpha_t$ and $\gamma_t$ as well as $c^S \mal I_t=\langle S \rangle_t$ that
\begin{align*}
E^Q[\varphi \mal \Delta S_T] &= E^Q\left[\frac{\widetilde{\Delta\varphi}^2-(\Delta\mathrm{NT})^2}{R} \mal \langle S \rangle_T \right]+O(\varepsilon).
\end{align*}
Again replacing -- at the leading order $O(\varepsilon^{2/3})$ -- $\widetilde{\Delta\varphi}_t^2$ by its expectation $(\Delta\mathrm{NT}_t)^2/3$ under the uniform distribution on $[-\Delta\mathrm{NT}_t,\Delta\mathrm{NT}_t]$ in the above time average, we obtain
\begin{align*}
E^Q[\varphi \mal \Delta S_T] &= -\frac{2}{3} E^Q\left[\frac{(\Delta\mathrm{NT})^2}{R} \mal \langle S \rangle_T \right]+O(\varepsilon).
\end{align*}

Finally, consider the second part $E^Q[\Delta\varphi \mal \Delta S_T]$ of the transaction cost loss. It can be computed by integrating the $Q$-drift rate of $\Delta\varphi \mal \Delta S_t$. As the diffusion coefficient of $\Delta S_t$ is of order $O(\varepsilon^{2/3})$ by \eqref{eq:ito}, Girsanov's theorem shows that the latter coincides with the corresponding drift under the physical probability at the leading order $O(\varepsilon^{2/3})$. In view of \eqref{eq:drift} and the definition of $\alpha_t$, the latter is $\frac{\Delta\varphi \widetilde{\Delta\varphi}}{R} \mal \langle S \rangle_t+O(\varepsilon)$. Now, insert $\Delta\varphi_t=\overline{\Delta\varphi}_t+\widetilde{\Delta\varphi}_t$ and once again approximate the oscillatory terms $\widetilde{\Delta\varphi}_t$ and $\widetilde{\Delta\varphi}_t^2$ by their expectations $0$ resp.\ $(\Delta\mathrm{NT}_t)^2/3$ under the uniform law on $[-\Delta\mathrm{NT}_t,\Delta\mathrm{NT}_t]$. Then:
$$E^Q[\Delta\varphi \mal \Delta S_T]=\frac{1}{3} E^Q\left[\frac{(\Delta\mathrm{NT})^2}{R} \mal \langle S \rangle_T \right]+O(\varepsilon).$$
As a consequence, the transaction cost loss $E^Q[(\varphi +\Delta\varphi) \mal \Delta S_T]+O(\varepsilon)$ is twice as large as the corresponding displacement loss at the leading order, and the total utility loss is given by 
$$U(x)-U^\varepsilon(x)=U'(x) E^Q\left[\frac{(\Delta\mathrm{NT})^2}{2R} \mal \langle S \rangle _T\right]+O(\varepsilon).$$
Formula \eqref{eq:celoss} for the certainty equivalent loss then follows by Taylor expansion.

\section{Derivation of the Implied Trading Volume}\label{sec:tv}

Next, we derive the formulas for the implied trading volume from Section~\ref{sec:impliedtv}. To this end, consider an arbitrary diffusion strategy $d\phi_t=b^\phi_t dt +\sqrt{c^\phi_t}dW_t$, implemented by performing the minimal amount of trading to keep the actual number $\phi^\varepsilon_t$ of risky shares in a symmetric buffer $[\phi_t-\Delta^\varepsilon_t,\phi_t+\Delta^\varepsilon_t]$ with halfwidth $\Delta^\varepsilon_t=O(\varepsilon^{\nu})$, $\nu>0$, around $\phi_t$. This means that trading of local time type occurs whenever the number $\phi_t^\varepsilon$ of risky shares reaches the moving boundaries $\phi_t \pm \Delta^\varepsilon_t$, and the actual number of risky shares is the difference between the cumulative numbers of shares purchased resp.\ sold: $\phi^\varepsilon_t=L_t-U_t$. Put differently, the difference $\Delta\phi_t=\phi^\varepsilon_t-\phi_t=L_t-U_t-\phi_t$ between the actual position $\phi^\varepsilon_t$ and the target $\phi_t$ is a diffusion reflected to remain in the small interval $[-\Delta^\varepsilon_t,\Delta^\varepsilon_t]$ with width of order $O(\varepsilon^\nu)$. To compute the turnover $||\phi^\varepsilon||_T=L_T+U_T$, we have to estimate the growth of the two local time processes, at the leading order for small $\varepsilon$.

To this end, rescale both time and space to obtain the rescaled process
$$(D_\tau)_{\tau \geq 0}=(\varepsilon^{-\nu} \Delta\phi_{\tau\varepsilon^{2\nu}})_{\tau \geq 0},$$
which has drift $b^D_\tau=\varepsilon^{-\nu} \varepsilon^{2\nu} b^\phi_{\tau\varepsilon^{2\nu}} =O(\varepsilon^{\nu})$ and squared diffusion coefficient $c^D_\tau=\varepsilon^{-2\nu}\varepsilon^{2\nu}c_{\tau\varepsilon^{2\nu}}=O(1)$ in the interior of the no-trade region, where $\phi^\varepsilon_t$ is constant. Now, divide the interval $[0,T]$ into a mesh $0=t^\varepsilon_0<\ldots<t^\varepsilon_{N^\varepsilon}=T$ with mesh size of order $O(\varepsilon^{\nu})$, set $\tau^\varepsilon_i=\varepsilon^{-2\nu}t_i^\varepsilon$, $i=0,\ldots,N^\varepsilon$, and decompose
$$\Delta\phi_T=\varepsilon^{\nu}\sum_{i=1}^{N^\varepsilon} (D_{\tau_i^\varepsilon}-D_{\tau^\varepsilon_{i-1}}).$$
As $\varepsilon$ becomes small, the time intervals $[\tau^\varepsilon_{i-1},\tau^\varepsilon_i]$ becomes longer and longer, whereas the reflecting barriers $\pm\varepsilon^{-\nu}\Delta^\varepsilon_{\tau\varepsilon^{2\nu}}$ remain approximately constant on $[\tau^\varepsilon_{i-1},\tau^\varepsilon_i]$. At the leading order, each increment $D_{\tau^\varepsilon_i}-D_{\tau^\varepsilon_{i-1}}$ therefore corresponds to the one of a driftless Brownian motion with variance $c^\phi_{\tau^\varepsilon_{i-1} 
\varepsilon^{2\nu}}$, reflected to remain in the interval $\varepsilon^{-\nu}[-\Delta^\varepsilon_{\tau^\varepsilon_{i-1}\varepsilon^{2\nu}},\Delta^\varepsilon_{\tau^\varepsilon_{i-1}\varepsilon^{2\nu}}]$. As the time interval $[\tau^\varepsilon_{i-1},\tau^\varepsilon_i]$ grows, the corresponding local times therefore approach the long-run averages for reflected Brownian motion, which have been derived, e.g., in \cite[Remark 4]{janecek.shreve.04}. As a result, cumulative purchases and sales coincide -- at the leading order $O(\varepsilon^{-\nu})$ for small transaction costs $\varepsilon_t=\varepsilon \mathcal{E}_t$ -- and are both given by
$$\varepsilon^{\nu} \sum_{i=1}^{N^\varepsilon} (\tau^\varepsilon_i-\tau^\varepsilon_{i-1}) \frac{c^{\phi}_{\tau^\varepsilon_{i-1}\varepsilon^{2\nu}}}{4 \varepsilon^{-\nu}\Delta^\varepsilon_{\tau^\varepsilon_{i-1}\varepsilon^{2\nu}}}=\sum_{i=1}^{N^\varepsilon} (t_i-t_{i-1}) \frac{c^\phi_{t_{i-1}}}{4\Delta^\varepsilon_{t_{i-1}}}.$$
As the mesh size becomes small, they therefore converge to the integrals
$$\int_0^T \frac{c^\phi_t}{4 \Delta^\varepsilon_t} dt = \int_0^T \frac{d\langle \phi\rangle_t}{4\Delta^\varepsilon_t}.$$
At the leading order $\varepsilon^{-\nu}$, the corresponding absolute turnover is thus given by
$$||\phi^\varepsilon||_T \sim  \int_0^T \frac{c^\phi_t}{2\Delta^\varepsilon_t}dt.$$
All these considerations hold for \emph{any} strategy that remains close to a diffusion by means of reflection off two symmetric moving boundaries. The explicit representation for the approximately optimal strategy $\varphi_t^{\varepsilon}$ from Section \ref{sec:policy} follows by inserting Formula \eqref{eq:deviation} for maximal deviations $\Delta^\varepsilon_t$, and taking into account that $\langle \varphi+\overline{\Delta\varphi} \rangle_t= \langle \varphi \rangle_t$ at the leading order $O(\varepsilon^{1/3})$.

\section{Mean-Variance Portfolio Selection}\label{sec:mvps}

In this section, we derive the results of Section \ref{ref:resultmvps} on mean-variance portfolio selection with small transaction costs. For the convenience of the reader, we first briefly recall the well-known frictionless case, and its connection to the maximization of (truncated) quadratic utility. 

\subsection{The Frictionless Case}\label{ss:mvps1}
The Markowitz portfolio selection problem of minimizing the portfolio's variance $\mathrm{Var}[X_T(\psi)]$ for a given mean $E[X_T(\psi)]=m>x$ is equivalent to minimizing $E[X_T^2(\psi)]-m^2-\lambda (E[X_T(\psi)]-m)$, for a Lagrange multiplier $\lambda \geq 0$ such that the constraint $E[X_T(\psi)]=m$ is satisfied. The optimal strategy in turn corresponds to the maximizer of $E[-(X_T(\psi)-\lambda/2)^2]$, that is, to the optimal strategy for the quadratic utility $u(x)=-x^2$ starting from initial endowment $x-\lambda/2$. By the homotheticity of the quadratic utility maximization problem, the latter is given by the $(\lambda/2-x)$-fold of the optimal strategy $\phi_t$ for quadratic utility and the standardized initial endowment $-1$. Setting up this portfolio only requires an initial endowment of $x-\lambda/2$; for the mean-variance efficient portfolio $\varphi_t$, the remaining endowment of $\lambda/2$ is kept in the safe asset. Hence, its wealth process is given by 
$$X_T(\varphi)=\lambda/2 + (\lambda/2-x)(-1+\phi \mal S_T).$$

Now, let $U(-1)=E[-(-1+\phi \mal S_T)^2]$ be the quadratic utility generated by the optimal portfolio for initial endowment $-1$. Then by, e.g., \cite[Lemma 3.1.5]{cerny.kallsen.07} the latter coincides with the mean of this portfolio: $U(-1)=E[-1+\phi \mal S_T]$. As a result, the Lagrange multiplier $\lambda$ is determined by the constraint $m=E[X_T(\varphi)]=\lambda/2 +(\lambda/2-x)U(-1)$ as
$$\frac{\lambda}{2}= \frac{m+ x U(-1)}{1+U(-1)}.$$
The corresponding minimal variance is in turn given by
\begin{align*}
\mathrm{Var}[X_T(\varphi)]&=\left(\frac{\lambda}{2}-x\right)^2 \left(E[(-1+\phi \mal S_T)^2]-E[-1+\phi \mal S_T]^2\right)\\
&=\left(\frac{m-x}{1+U(-1)}\right)^2 (-U(-1)-U(-1)^2)=(m-x)^2 \frac{-U(-1)}{1+U(-1)}.
\end{align*}
As a result, for any target mean $m>x$, the corresponding optimal Sharpe ratio is the same:
$$\mathrm{SR}= \frac{E[X_T(\varphi)]-x}{\sqrt{\mathrm{Var}[X_T(\varphi)]}}=\sqrt{-\frac{1}{U(-1)}-1}.$$
In particular, the maximal return for a given variance bound $s^2$ is $E[X_T(\varphi)-x]=s\mathrm{SR}$.

\subsection{Small Transaction Costs}\label{ss:mvps2}

Let us now consider how the above results adapt to small proportional transaction costs $\varepsilon_t=\varepsilon \mathcal{E}_t$. Maximizing expected quadratic utility is covered by the results of the previous sections, except for the fact that $u(x)=-x^2$ is not increasing for wealth levels beyond the bliss point $x=0$. However, for continuous asset prices, the optimal portfolio starting from initial wealth $-1$ never crosses the bliss point, cf., e.g., \cite[Lemma 3.7]{cerny.kallsen.07}; thereby, it is also optimal for the truncated quadratic utility $-\min(x,0)^2$, which fits into our setting. Hence, the optimal strategy for the shadow price from Appendix D is also optimal with transaction costs. Indeed, it is optimal for quadratic utility and thereby also for the truncated version in the shadow market. As the latter provides better terms of trade for any portfolio, but the same for the optimizer, the corresponding optimal portfolio is also optimal for the \emph{monotone} truncated quadratic utility in the original market with transaction costs. But as the portfolio always remains below the bliss point $x=0$, it is a fortiori also optimal for the quadratic utility with transaction costs. 

By the homotheticity of the quadratic utility maximization problem, the risk-tolerance process for $u(x)=-x^2$ is given by 
$$R_t=1-\phi \mal S_t.$$
In view of (3.3), the leading-order optimal strategy for quadratic utility and initial endowment $-1$ with transaction costs $\varepsilon_t$ therefore keeps the number of shares $\phi^\varepsilon_t$ within a no-trade region centered around the midpoint $\overline{\mathrm{NT}}^\phi_t=\phi_t+\phi'_t (X^\varepsilon_t(\phi^\varepsilon)-X_t(\phi))=\phi_t X^\varepsilon_t(\phi^\varepsilon)/X_t(\phi)$,\footnote{Here, $X^\varepsilon_t(\phi^\varepsilon)$ and $X_t(\phi)$ dentote the wealth processes generated by trading $\phi^\varepsilon_t$ and $\phi_t$, respectively, starting from initial endowment $-1$.} with halfwidth
$$\Delta\mathrm{NT}^\phi_t=\left(\frac{3\varepsilon_t(1-\phi \mal S_t)}{2} \frac{d\langle\phi \rangle_t}{d\langle S\rangle_t}\right)^{1/3}.$$
The mean-variance optimal portfolio $\varphi^\varepsilon_t$ satisfying the constraint $E[X_T^\varepsilon(\varphi^\varepsilon)]=m$ is obtained by holding a cash position of $\lambda^\varepsilon/2$ and trading the  $(\lambda^\varepsilon/2-x)$-fold of $\phi^\varepsilon$, where $\lambda^\varepsilon$ refers to the Lagrange multiplier $\lambda^\varepsilon=2(m+xU^\varepsilon(-1))/(1+U^\varepsilon(-1))$ for the shadow price $S^\varepsilon_t$. As a result, the no-trade region for $\varphi^\varepsilon_t$ is obtained by simply rescaling the one for $\phi^\varepsilon_t$ by a factor of $(\lambda^\varepsilon/2-x)$. 

Let us now turn to the corresponding welfare effect of small transaction costs. For the quadratic utility $u(x)=-x^2$ and initial endowment $-1$, the latter is determined by Formula (3.4):
\begin{align*}
 U^\varepsilon(-1) &\sim U\left(-1-E^Q\left[\int_0^T \frac{(\Delta\mathrm{NT}^\phi_t)^2}{2(1-\phi \mal S_t)}d\langle S \rangle_t\right]\right) 
\sim U(-1)\left(1+2E^Q\left[\int_0^T \frac{(\Delta\mathrm{NT}^\phi_t)^2}{2(1-\phi \mal S_t)}d\langle S \rangle_t\right]\right),
\end{align*}
 where the second step follows from the homotheticity of the quadratic utility maximization problem and Taylor expansion. As a result, the maximal Sharpe ratios in the presence of small transaction costs $\varepsilon_t$ are given by
 \begin{equation}\label{eq:Sharpetotal}
 \mathrm{SR}^\varepsilon_t = \sqrt{-\frac{1}{U^\varepsilon(-1)}-1} \sim \mathrm{SR}-\frac{1+\mathrm{SR}^2}{\mathrm{SR}} E^Q\left[\int_0^T \frac{(\Delta\mathrm{NT}^\phi_t)^2}{2(1-\phi \mal S_t)}d\langle S \rangle_t\right].
 \end{equation}
For a given target mean $m$, the corresponding minimal variance is increased due to small transaction costs to
$$\mathrm{Var}[X^\varepsilon_T(\varphi^\varepsilon)]=(m-x)^2\frac{-U^\varepsilon(-1)}{1+U^\varepsilon(-1)} \sim \mathrm{Var}[X_T(\varphi)]\left(1+\frac{2}{1+U(-1)}E^Q\left[\int_0^T \frac{(\Delta\mathrm{NT}^\phi_t)^2}{2(1-\phi \mal S_t)}d\langle S \rangle_t\right]\right).$$
Conversely, small transaction costs reduce the maximal expected return for a given variance bound $s^2$ from $s \mathrm{SR}$ to
$$s\mathrm{SR}^\varepsilon \sim s \left(\mathrm{SR} -\frac{1+\mathrm{SR}^2}{\mathrm{SR}} E^Q\left[\int_0^T \frac{(\Delta\mathrm{NT}^\phi_t)^2}{2(1-\phi \mal S_t)}d\langle S \rangle_t\right]\right).$$

Now, consider the composition of the Sharpe ratio loss caused by small transaction costs. To this end, estimate the Sharpe ratio of the frictional optimizer $\varphi^\varepsilon_t$ traded at the frictionless mid price $S_t$. First, notice that the Sharpe ratio is invariant among different multiples of the optimal strategy $\phi^\varepsilon_t$ for quadratic utility:
\begin{equation}\label{eq:Sharpe}
\frac{E[x+\theta \phi^\varepsilon \mal S_T]-x}{\sqrt{\mathrm{Var}[x+\theta \phi^\varepsilon \mal S_T]}}=\frac{E[\phi^\varepsilon \mal S_T]}{\sqrt{\mathrm{Var}[\phi^\varepsilon \mal S_T]}} \quad \mbox{for any $\theta \in (0,\infty)$.}
\end{equation}
In contrast, the corresponding expected quadratic utility $E[-(-1+\theta\phi^\varepsilon \mal S_T)^2]$ depends on the multiplier and is maximized for $\theta=E[\phi^\varepsilon \mal S_T]/E[(\phi^\varepsilon \mal S_T)^2]$. Then, $E[-1+\theta\phi^\varepsilon \mal S_T]=E[-(-1+\theta\phi^\varepsilon \mal S_T)^2]$, so that the corresponding Sharpe ratio is given by
$$\frac{E[x+\theta \phi^\varepsilon \mal S_T]-x}{\sqrt{\mathrm{Var}[x+\theta \phi^\varepsilon \mal S_T]}}=\sqrt{\frac{1}{E[(-1+\theta\phi^\varepsilon \mal S_T)^2]}-1}.$$
By \eqref{eq:Sharpe}, this Sharpe ratio coincides with the one for the frictional optimizer $\varphi^\varepsilon_t$, as the latter is also obtained from $\phi^\varepsilon_t$ by rescaling. As the quadratic utility derived from $\varphi^\varepsilon_t$ is -- by definition of $\theta$ -- smaller than the one for $\theta\phi^\varepsilon_t$, this implies that
\begin{equation}\label{eq:Sharpeest}
\frac{E[x+ \varphi^\varepsilon \mal S_T]-x}{\sqrt{\mathrm{Var}[x+ \varphi^\varepsilon \mal S_T]}}=\sqrt{\frac{1}{E[(-1+\theta\phi^\varepsilon \mal S_T)^2]}-1} \geq \sqrt{\frac{1}{E[(-1+\varphi^\varepsilon \mal S_T)^2]}-1}.
\end{equation}
For the quadratic utility $u(x)=-x^2$, the computations in Appendix \ref{sec:uloss} show that the displacement loss, i.e., the difference between the optimal frictionless utility $E[-(-1+\phi \mal S_T)^2]$ and the performance $E[-(-1+\phi^\varepsilon \mal S_T)^2]$ of the frictional optimizer traded at the mid price, amounts to one third of the total utility loss at the leading order. Together with \eqref{eq:Sharpeest} and Taylor expansion, this shows that the reduction of the Sharpe ratio due to displacement amounts to at most one third of the total Sharpe ratio loss in \eqref{eq:Sharpetotal}.

\section{Long-Run Growth Optimality}\label{|sec:numeraire}
In this section, we argue why -- without random endowment ($\Psi_t=0$) -- our approximate log-optimal portfolio not only maximizes the \emph{expected} long-term growth rate, but also its almost sure counterpart \eqref{eq:growth}, at the leading order for small costs. In addition, we also compute the leading-order reduction \eqref{eq:growthrate} of the growth rate due to small transaction costs. To this end, we proceed similarly as in Appendices \ref{sec:opt} and \ref{sec:uloss}, but work directly in terms of relative quantities to deal with the double limit of small transaction costs and a long horizon. Recall that $\eta_t=\varepsilon_t/S_t=\varepsilon \mathcal{E}_t/S_t$ denotes the relative bid-ask spread, set
 \begin{align*}
\alpha_t =\frac{c^Y_t}{3 c^{\pi(1-\pi)\mal Y-\pi}_t}, \quad \Delta\pi_t=\left(\frac{3 \eta_t}{2} \frac{c^{\pi(1-\pi)\mal Y-\pi}_t}{c^Y_t}\right)^{1/3}, \quad \gamma_t=3\alpha_t (\Delta\pi_t)^2,
\end{align*}
and let $\pi^\varepsilon_t$ be the risky weight corresponding to the minimal amount of trading necessary to remain in the no-trade region $[\pi_t-\Delta\pi_t,\pi_t+\Delta\pi_t]$ around the frictionless target $\pi_t$. This means that the corresponding number $\varphi^\varepsilon_t$ of risky shares is constant while the deviation $\widetilde{\Delta\pi_t}=\pi^\varepsilon_t-\pi_t$ lies in $(\-\Delta\pi_t,\Delta\pi_t)$. With this notation, define
$$\Delta Y_t= \alpha_t \widetilde{\Delta\pi}_t^3 - \gamma_t \widetilde{\Delta\pi}_t.$$
Then by definition of $\alpha_t,\gamma_t,$ and $\Delta\pi_t$, one readily verifies that $\Delta Y_t$ decreases from $\eta_t$ to $-\eta_t$ as $\widetilde{\Delta\pi}_t$ moves from $-\Delta\pi_t$ to $\Delta \pi_t$. Thus,
$$S^\varepsilon_t= S_t(1+\Delta Y_t)$$
is a valid candidate shadow price process, in that it takes values in the bid-ask spread and coincides with the ask resp.\ bid price whenever purchases resp.\  sales occur for the policy $\pi^\varepsilon_t$. In particular, the frictional wealth process corresponding to the risky weight $\pi^\varepsilon_t$ coincides with its frictionless counterpart for $S^\varepsilon_t$ and is therefore given by the stochastic exponential $x\E(\frac{\pi^\varepsilon}{S^\varepsilon} \mal S^\varepsilon)_t$. As the ratio $S_t/S_t^\varepsilon$ is of the form $1+O(\varepsilon)$, the deviation $\Delta Y_t$ is of order $O(\varepsilon)$, and $S_t=S_0 \E(Y)_t=S_0+S \mal Y_t$, the definition of $S^\varepsilon_t$ and integration by parts yield
\begin{align*}
\frac{\pi^\varepsilon}{S^\varepsilon} \mal S^\varepsilon_t&= \frac{\pi^\varepsilon}{S^\varepsilon} \mal \left(S_t+\Delta Y \mal S_t +S \mal \Delta Y_t + \langle S, \Delta Y \rangle_t\right)\\
&= \pi^\varepsilon \mal Y_t +\pi^\varepsilon \frac{S}{S^\varepsilon} \mal (\Delta Y_t + \langle Y, \Delta Y \rangle_t)=\pi^\varepsilon \mal (Y_t+\Delta Y_t + \langle Y, \Delta Y \rangle_t)+O(\varepsilon T).
\end{align*}
As a result, the frictional wealth process corresponding to $\pi^\varepsilon_t$ is given by\footnote{As we exclusively use relative quantities in this appendix, the arguments of wealth processes refer to risky weights rather than numbers of risky shares here.}
\begin{equation}\label{eq:wealth}
X^\varepsilon_t(\pi^\varepsilon)=x\E\Big(\pi^\varepsilon \mal \big(Y+\Delta Y+\langle Y, \Delta Y \rangle \big)\Big)_t \times e^{O(\varepsilon T)}.
\end{equation}
For an arbitrary risky weight $\vartheta_t$, the corresponding expression provides an upper bound for the frictional wealth, because trades in the shadow market take place at potentially more favorable prices. As a result, for any risky weight $\vartheta^\varepsilon_t$, the ratio of wealth processes satisfies
\begin{align*}
\frac{X_T^\varepsilon(\vartheta^\varepsilon)}{X^\varepsilon_T(\pi^\varepsilon)} & \leq \frac{\E(\vartheta^\varepsilon \mal (Y+\Delta Y+\langle Y, \Delta Y \rangle))_T}{\E(\pi^\varepsilon \mal (Y+\Delta Y+\langle Y, \Delta Y \rangle))_T}\times e^{O(\varepsilon T)}\\
&=\E\Big((\vartheta^\varepsilon-\pi^\varepsilon) \mal \big(Y+\Delta Y+\langle Y,\Delta Y \rangle- \pi^\varepsilon \mal \langle Y+\Delta Y\rangle\big)\Big) \times e^{O(\varepsilon T)},
\end{align*}
where the second equality follows from Yor's formula as in \cite[Lemma 3.4]{karatzas.kardaras.07}. Like in the general case, it suffices to consider families of competitors $(\vartheta^\varepsilon_t)_{\varepsilon>0}$ converging to the frictionless optimizer $\pi_t$ as $\varepsilon \downarrow 0$, so that $\vartheta^\varepsilon_t-\pi^\varepsilon_t=o(1)$.

We now show that, by the respective definitions of $\alpha_t$, $\gamma_t$, and $\Delta\pi_t$, the process $Y_t+\Delta Y_t+\langle Y,\Delta Y \rangle_t-  \pi^\varepsilon \mal\langle Y+\Delta Y\rangle_t$ is a martingale up to a drift rate of order $O(\varepsilon^{2/3})$, so that the ratio of wealth processes is bounded from above by a martingale up to a multiplicative finite variation process $D^\varepsilon_t$ of order $e^{o(\varepsilon^{2/3}T)}$. To this end, we first compute the dynamics of $\Delta Y_t$. Integration by parts shows that the risky weight $\pi^\varepsilon_t$ has dynamics
\begin{align*}
\pi^\varepsilon_t=\frac{\varphi^\varepsilon_t S^\varepsilon_t}{X^\varepsilon_t(\pi^\varepsilon)}&=\varphi^\varepsilon_t \E((1-\pi^\varepsilon)\mal Y+ (1-\pi^\varepsilon)S/S^\varepsilon \mal \Delta Y + \text{finite variation terms})_t\\
&= \pi^\varepsilon(1-\pi^\varepsilon) \mal Y_t+ \pi^\varepsilon(1-\pi^\varepsilon)S/S^\varepsilon \mal \Delta Y_t + \text{finite variation terms},
\end{align*}
because the frictional number of risky shares $\varphi^\varepsilon_t$ is of finite variation. As a result, the martingale part of $\widetilde{\Delta\pi}_t=\pi^\varepsilon_t-\pi_t$ matches the one of $\pi^\varepsilon(1-\pi^\varepsilon) \mal Y_t+ \pi^\varepsilon(1-\pi^\varepsilon)S/S^\varepsilon \mal\Delta Y_t-\pi_t$. Hence, using integration by parts and It\^o's formula to write
$$\Delta Y_t= 3\alpha \widetilde{\Delta\pi} \mal \langle \widetilde{\Delta\pi} \rangle_t +(3\alpha\widetilde{\Delta\pi}^2-\gamma) \mal \widetilde{\Delta\pi}_t +\widetilde{\Delta\pi}^3 \mal \alpha_t -\widetilde{\Delta\pi} \mal \gamma_t +\langle 3\widetilde{\Delta\pi}^2 \mal \alpha -\gamma, \widetilde{\Delta\pi} \rangle_t$$
shows that the diffusion coefficient of $\Delta Y_t$ is given by the $(3\alpha_t\widetilde{\Delta\pi}_t^2-\gamma_t)$-fold of its counterpart for $\pi(1-\pi)\mal Y-\pi$, up to terms of order $O(\varepsilon)$. In particular, it is of order $O(\varepsilon^{2/3})$. The drift rate of $\Delta Y_t$ is in turn given by $b^{\Delta Y}_t= 3\alpha_t \widetilde{\Delta\pi}_t c^{\widetilde{\Delta \pi}}_t+O(\varepsilon^{2/3})=3\alpha_t \widetilde{\Delta\pi}_t c^{\pi(1-\pi)\mal Y-\pi}_t+O(\varepsilon^{2/3})$. As a result, the drift rate of the process $Y_t+\Delta Y_t+\langle Y,\Delta Y \rangle_t-  \pi^\varepsilon \mal\langle Y+\Delta Y\rangle_t$ indeed vanishes at the leading order:
$$b^Y_t+b^{\Delta Y}_t+c^{Y \Delta Y}_t -(\pi_t+\widetilde{\Delta\pi}_t) (c^Y_t+2c^{Y \Delta Y}_t+c^{\Delta Y}_t)=3\alpha_t \widetilde{\Delta\pi}_t c^{\pi(1-\pi)\mal Y-\pi}_t-\widetilde{\Delta\pi}_t c^Y_t+O(\varepsilon^{2/3})=O(\varepsilon^{2/3}),$$ 
by definition of $\alpha_t$ and because the frictionless growth-optimal portfolio is given by $\pi_t=b^Y_t/c^Y_t$. In summary, $X^\varepsilon_t(\vartheta^\varepsilon)/X^\varepsilon_t(\pi^\varepsilon) \leq D^\varepsilon_t M^\varepsilon_t$ for a martingale $M^\varepsilon_t$ and a finite variation process $D^\varepsilon_t$, both positive and starting at $1$, and satisfying
\begin{equation}\label{eq:D}
\lim_{T \to \infty} \frac{1}{T} \log D^\varepsilon_T =o(\varepsilon^{2/3}).
\end{equation}

Let us now argue why this implies the growth-optimality of the proposed policy, at the leading order for small costs. Here, the argument follows its frictionless counterpart \cite[Theorem 3.10.1]{karatzas.shreve.98}, up to taking care of the remainder terms in an appropriate manner. Fix $\delta \in (0,1)$; then, Doob's maximal inequality (e.g., \cite[Corollary 4.8 and Theorem 4.2]{elliott.82}) implies
$$e^{\delta n} P\left[\sup_{t \in [n,\infty)}\frac{X_t^\varepsilon(\vartheta^\varepsilon)}{X_t^\varepsilon(\pi^\varepsilon)D^\varepsilon_t} > e^{\delta n}\right] \leq e^{\delta n} P\left[\sup_{t \in [n,\infty)}M^\varepsilon_t > e^{\delta n}\right] \leq  E[M^\varepsilon_0]=1,$$
for all $n \in \mathbb{N}$. As a consequence:
$$\sum_{n=1}^\infty P\left[\sup_{t \in [n,\infty)} \frac{1}{n} \log \frac{X_t^\varepsilon(\vartheta^\varepsilon)}{X_T^\varepsilon(\pi^\varepsilon)D^\varepsilon_t} > \delta \right] \leq \sum_{n=1}^\infty e^{-\delta n} < \infty.$$
In view of the Borel-Cantelli lemma, this shows that $P$-a.s.\ there exists some (random) $n_0 \in \mathbb{N}$ such that for all $n \geq n_0$ we have $\sup_{t \in [n,\infty)}  \frac{1}{n} \log \frac{X_t^\varepsilon(\vartheta^\varepsilon)}{X_T^\varepsilon(\pi^\varepsilon)D^\varepsilon_t} \leq \delta$ and hence
$$\sup_{t \in [n,\infty)} \frac{1}{t} \log \frac{X_t^\varepsilon(\vartheta^\varepsilon)}{X_T^\varepsilon(\pi^\varepsilon)D^\varepsilon_t} \leq \delta.$$ 
This in turn yields that, $P$-a.s.,
$$\limsup_{T \to \infty} \frac{1}{T} \log X^\varepsilon_T(\vartheta^\varepsilon) \leq  \limsup_{T \to \infty} \frac{1}{T} \log X^\varepsilon_T(\pi^\varepsilon)+\delta +\limsup_{T \to \infty} \frac{1}{T} \log D^\varepsilon_T.$$
As $\delta$ was arbitrary, combining this with \eqref{eq:D} shows that the risky weight $\pi^\varepsilon_t$ is indeed growth optimal at the leading order $O(\varepsilon^{2/3})$.

\medskip

Let us now compute the leading-order reduction of the maximal long-run growth rate due to the presence of small transaction costs. To this end, consider the log-ratio of the optimal frictional and frictionless wealth processes. In view of \eqref{eq:wealth}, it can be written as
\begin{align}
&\frac{1}{T}\log \frac{X^\varepsilon_T(\pi^\varepsilon)}{X_T(\pi)}\notag\\
\qquad&= \frac{1}{T}\log \frac{ \E(\pi^\varepsilon \mal (Y +\Delta Y +\langle Y, \Delta Y \rangle))_T}{\E(\pi \mal Y)_T}+O(\varepsilon)\notag \\
\qquad &= \frac{1}{T}\Big( \pi^\varepsilon \mal (Y_T+\Delta Y_T +\langle Y, \Delta Y \rangle_T) -\frac{(\pi^\varepsilon)^2}{2} \mal \langle  Y+\Delta Y \rangle_T -\pi \mal Y_T +\frac{\pi^2}{2} \mal \langle Y \rangle_T\Big)+O(\varepsilon)\notag \\
\qquad &= \frac{1}{T}\Big(\widetilde{\Delta\pi} \mal (Y_T - \pi \mal \langle Y \rangle_T)+\pi \mal \Delta Y_T -\frac{\widetilde{\Delta\pi}^2}{2} \mal \langle Y \rangle_T +\widetilde{\Delta\pi} \mal \Delta Y_T \notag\\
\qquad & \qquad \qquad  +\pi(1-\pi) \mal \langle \Delta Y, Y \rangle_T\Big) +O(\varepsilon).\label{eq:goal}
\end{align}
Here, we have used in the last step that the diffusion coefficient of $\Delta Y_t$ is of order $O(\varepsilon^{2/3})$. The first term on the right-hand side of \eqref{eq:goal} is a martingale: the drift rate of $Y_t-\pi \mal \langle Y \rangle_t$ is given by $b^Y_T-(b^Y_t/c^Y_t)c^Y_t=0$. In view of the Dambis-Dubins-Schwarz theorem, it can therefore be written as a time-changed Brownian motion. Hence, its long-term average vanishes by the law of the iterated logarithm, provided that the local variance $c^Y_t$ of the returns is not too far from stationary:
\begin{equation}\label{est:1}
\limsup_{T \to \infty} \frac{1}{T} \left( \widetilde{\Delta\pi} \mal (Y_T - \pi \mal \langle Y \rangle_T)\right)=0.
\end{equation}
As for the next term, recall $\Delta Y_t=O(\varepsilon)$. Then, integration by parts yields
\begin{align*}
\frac{1}{T} \left(\pi \mal \Delta Y_T\right)&=\frac{1}{T}\Big( \pi_T \Delta Y_T -\pi_0 \Delta Y_0 -\Delta Y \mal \pi_T -\langle \Delta Y, \pi \rangle_T\Big)\\
&= -\frac{1}{T}\langle \Delta Y, \pi \rangle_T+O(\varepsilon)= \frac{1}{T}\left(3\alpha(\Delta\pi^2-\widetilde{\Delta\pi}^2)c^{\pi(1-\pi)\mal Y-\pi,\pi} \right)\mal I_T +O(\varepsilon),
\end{align*}
where we have used for the last step that the martingale part of $\Delta Y_t$ coincides with the one of $(3\alpha\widetilde{\Delta\pi}^2-\gamma)\mal(\pi(1-\pi)\mal Y_t-\pi_t)$ at the order $O(\varepsilon^{2/3})$, and have inserted the definition of $\gamma_t$. As in Appendices~\ref{sec:uloss} and \ref{sec:tv}, now approximate -- at the leading order -- the oscillatory deviation $\widetilde{\Delta\pi}_t^2$ by its expectation $\Delta\pi_t^2/3$ under the uniform law on $[-\Delta\pi_t,\Delta\pi_t]$ in the above time average, obtaining
\begin{equation}\label{est:2}
\frac{1}{T} \left(\pi \mal \Delta Y_T\right)= \frac{1}{T}\left(\alpha \Delta\pi^2 c^{\pi(1-\pi)\mal Y-\pi, 2\pi} \mal I_T\right) +O(\varepsilon).
\end{equation}
For the third term on the right-hand side of \eqref{eq:goal} we use the same averaging argument:
\begin{equation}\label{est:3}
-\frac{1}{T} \left(\frac{\widetilde{\Delta\pi}^2}{2} \mal \langle Y \rangle_T\right)= -\frac{1}{T} \left(\frac{\Delta\pi^2}{6} c^Y \mal I_T\right) +O(\varepsilon).
\end{equation}
Next, consider the fourth term on the right-hand side of \eqref{eq:goal}. As before, its martingale part does not contribute to the corresponding long-term average. For its drift part, inserting the representation determined above gives
$$\frac{1}{T} \left(\widetilde{\Delta\pi} b^{\Delta Y} \mal I_T\right)= \frac{1}{T} \left(3\alpha \widetilde{\Delta\pi}^2 c^{\pi(1-\pi)\mal Y-\pi} \mal I_T\right)+O(\varepsilon)=\frac{1}{T} \left(\alpha \Delta\pi^2 c^{\pi(1-\pi)\mal Y-\pi} \mal I_T\right)+O(\varepsilon),$$
where we have again approximated $\widetilde{\Delta\pi}_t^2$ by its expectation $\Delta\pi_t^2/3$ under the uniform law on $[-\Delta\pi_t,\Delta\pi_t]$. In summary,
\begin{equation}\label{est:4}
\limsup_{T \to \infty} \frac{1}{T} \left(\widetilde{\Delta\pi} \mal \Delta Y_T\right)= \limsup_{T \to \infty} \frac{1}{T} \left(\alpha \Delta\pi^2 c^{\pi(1-\pi)\mal Y-\pi} \mal I_T\right)+O(\varepsilon).
\end{equation}
Finally, let us turn to the last term on the right-hand side of \eqref{eq:goal}. Inserting the leading-order martingale part of $\Delta Y_t$, it follows that 
\begin{align}
\frac{1}{T}\left(\pi(1-\pi) \mal \langle \Delta Y, Y \rangle_T\right) &= \frac{1}{T}\left((3\alpha \widetilde{\Delta\pi}^2-\gamma) c^{\pi(1-\pi)\mal Y-\pi, \pi(1-\pi)\mal Y} \mal I_T\right)+O(\varepsilon) \notag\\
&= \frac{1}{T} \left( \alpha \Delta\pi^2  c^{\pi(1-\pi)\mal Y-\pi, -2\pi(1-\pi)\mal Y} \mal I_T\right)+O(\varepsilon).\label{est:5}
\end{align}
Here, we have used the definition of $\gamma_t$ in the second step and also applied the above averaging argument one more time. Now, inserting (\ref{est:1}--\ref{est:5}) and the definition of $\alpha_t$ into \eqref{eq:goal} gives 
\begin{align*}
\limsup_{T \to \infty} \frac{1}{T} \frac{X^\varepsilon_T(\pi^\varepsilon)}{X_T(\pi)}&=\limsup_{T \to \infty}  \frac{1}{T}\left(\left(-\alpha c^{\pi(1-\pi)\mal Y-\pi} -\frac{c^Y}{6}\right) \Delta\pi^2 \mal I_T\right)+O(\varepsilon)\\
&= \limsup_{T \to \infty} \left(-\frac{1}{T} \frac{\Delta\pi^2}{2} \mal \langle Y \rangle_T\right)+O(\varepsilon).
\end{align*}
Provided all limits exist, this yields the desired formula for the reduction of the long-term growth rate caused by the presence of small transaction costs. 

\bibliographystyle{abbrv}
\bibliography{tractrans}

\begin{thebibliography}{10}

\bibitem{acharya.pedersen.05}
V.~Acharya and L.~Pedersen.
\newblock Asset pricing with liquidity risk.
\newblock {\em J. Financ. Econ.}, 77(2):375--410, 2005.

\bibitem{almgren.03}
R.~Almgren.
\newblock Optimal execution with nonlinear impact functions and
  trading-enhanced risk.
\newblock {\em Appl. Math. Finance}, 10(1):1--18, 2003.

\bibitem{balduzzi.lynch.99}
P.~Balduzzi and A.~Lynch.
\newblock Transaction costs and predictability: some utility cost calculations.
\newblock {\em J. Financ. Econom.}, 52(1):47--78, 1999.

\bibitem{bichuch.11}
M.~Bichuch.
\newblock Asymptotic analysis for optimal investment in finite time with
  transaction costs.
\newblock {\em SIAM J. Financial Math.}, 3(1):433--458, 2011.

\bibitem{bjoerk.03}
T.~Bj\"ork.
\newblock {\em Arbitrage Theory in Continuous Time}.
\newblock Oxford University Press, second edition, 2003.

\bibitem{breiman.60}
L.~Breiman.
\newblock Investment policies for expanding businesses optimal in the long run.
\newblock {\em Nav. Res. Log.}, 7(4):647--651, 1960.

\bibitem{cerny.kallsen.07}
A.~{\v{C}}ern{\'y} and J.~Kallsen.
\newblock On the structure of general mean-variance hedging strategies.
\newblock {\em Ann. Probab.}, 35(4):1479--1531, 2007.

\bibitem{chacko.viceira.05}
G.~Chacko and L.~Viceira.
\newblock Dynamic consumption and portfolio choice with stochastic volatility
  in incomplete markets.
\newblock {\em Rev. Finan. Stud.}, 18(4):1369--1402, 2005.

\bibitem{dufresne.al.12}
P.~Collin-Dufresne, K.~Daniel, C.~Moallemi, and M.~Saglam.
\newblock Strategic asset allocation with predictable returns and transaction
  costs.
\newblock Preprint, 2012.

\bibitem{constantinides.86}
G.~Constantinides.
\newblock {Capital market equilibrium with transaction costs}.
\newblock {\em J. Polit. Econ.}, 94(4):842--862, 1986.

\bibitem{cvitanic.karatzas.96}
J.~Cvitani{\'c} and I.~Karatzas.
\newblock Hedging and portfolio optimization under transaction costs: a
  martingale approach.
\newblock {\em Math. Finance}, 6(2):133--165, 1996.

\bibitem{dai.al.10}
M.~Dai, Z.~Q. Xu, and X.~Y. Zhou.
\newblock Continuous-time {M}arkowitz's model with transaction costs.
\newblock {\em SIAM J. Financial Math.}, 1(1):96--125, 2010.

\bibitem{davis.97}
M.~H.~A. Davis.
\newblock Option pricing in incomplete markets.
\newblock In {\em Mathematics of Derivative Securities}, pages 216--226.
  Cambridge University Press, Cambridge, 1997.

\bibitem{davis.norman.90}
M.~H.~A. Davis and A.~R. Norman.
\newblock Portfolio selection with transaction costs.
\newblock {\em Math. Oper. Res.}, 15(4):676--713, 1990.

\bibitem{dumas.luciano.91}
B.~Dumas and E.~Luciano.
\newblock {An exact solution to a dynamic portfolio choice problem under
  transaction costs}.
\newblock {\em J. Finance}, 46(2):577--595, 1991.

\bibitem{elliott.82}
R.~J. Elliott.
\newblock {\em Stochastic Calculus and Applications}.
\newblock Springer, New York, 1982.

\bibitem{epps.76}
T.~Epps.
\newblock The demand for brokers' services: the relation between security
  trading volume and transaction cost.
\newblock {\em Bell J. Econ.}, 7(1):163--194, 1976.

\bibitem{garleanu.pedersen.12}
N.~Garleanu and L.~Pedersen.
\newblock Dynamic trading with predictable returns and transaction costs.
\newblock {\em J. Finance}, 68(6):2309--2340, 2013.

\bibitem{gerhold.al.11}
S.~Gerhold, P.~Guasoni, J.~Muhle-Karbe, and W.~Schachermayer.
\newblock {Transaction costs, trading volume, and the liquidity premium}.
\newblock {\em Finance Stoch.}, 18(1):1--37, 2014.

\bibitem{goodman.ostrov.10}
J.~Goodman and D.~N. Ostrov.
\newblock Balancing small transaction costs with loss of optimal allocation in
  dynamic stock trading strategies.
\newblock {\em SIAM J. Appl. Math.}, 70(6):1977--1998, 2010.

\bibitem{guasoni.robertson.12}
P.~Guasoni and S.~Robertson.
\newblock Portfolios and risk premia for the long run.
\newblock {\em Ann. Appl. Probab.}, 22(1):239--284, 2012.

\bibitem{guasoni.weber.12}
P.~Guasoni and M.~Weber.
\newblock Dynamic trading volume.
\newblock {\em Math. Finance}, to appear, 2015.

\bibitem{janecek.shreve.04}
K.~Jane{\v{c}}ek and S.~E. Shreve.
\newblock Asymptotic analysis for optimal investment and consumption with
  transaction costs.
\newblock {\em Finance Stoch.}, 8(2):181--206, 2004.

\bibitem{janecek.shreve.10}
K.~Jane{\v{c}}ek and S.~E. Shreve.
\newblock Futures trading with transaction costs.
\newblock {\em Illinois J. Math.}, 54(4):1239--1284, 2010.

\bibitem{liu.loewenstein.07}
B.~Jang, H.~Koo, H.~Liu, and M.~Loewenstein.
\newblock Liquidity premia and transaction costs.
\newblock {\em J. Finance}, 62(5):2329--2366, 2007.

\bibitem{kallsen.muhlekarbe.10}
J.~Kallsen and J.~Muhle-Karbe.
\newblock On using shadow prices in portfolio optimization with transaction
  costs.
\newblock {\em Ann. Appl. Probab.}, 20(4):1341--1358, 2010.

\bibitem{kallsen.muhlekarbe.12}
J.~Kallsen and J.~Muhle-Karbe.
\newblock Option pricing and hedging with small transaction costs.
\newblock {\em Math. Finance}, to appear, 2012.

\bibitem{karatzas.kardaras.07}
I.~Karatzas and C.~Kardaras.
\newblock The num\'eraire portfolio in semimartingale financial models.
\newblock {\em Finance Stoch.}, 11(4):447--493, 2007.

\bibitem{karatzas.shreve.98}
I.~Karatzas and S.~E. Shreve.
\newblock {\em Methods of Mathematical Finance}.
\newblock Springer, New York, 1998.

\bibitem{karatzas.zitkovic.03}
I.~Karatzas and G.~{\v{Z}}itkovi{\'c}.
\newblock Optimal consumption from investment and random endowment in
  incomplete semimartingale markets.
\newblock {\em Ann. Probab.}, 31(4):1821--1858, 2003.

\bibitem{karpoff.87}
J.~Karpoff.
\newblock The relation between price changes and trading volume: a survey.
\newblock {\em J Financ. Quant. Anal.}, 22(1):109--126, 1987.

\bibitem{kelly.56}
J.~Kelly.
\newblock A new interpretation of information rate.
\newblock {\em AT\&T Tech J.}, 35:917--926, 1956.

\bibitem{kim.omberg.96}
T.~Kim and E.~Omberg.
\newblock Dynamic nonmyopic portfolio behavior.
\newblock {\em Rev. Finan. Stud.}, 9(1):141--161, 1996.

\bibitem{kramkov.sirbu.06a}
D.~Kramkov and M.~S{\^{\i}}rbu.
\newblock On the two-times differentiability of the value functions in the
  problem of optimal investment in incomplete markets.
\newblock {\em Ann. Appl. Probab.}, 16(3):1352--1384, 2006.

\bibitem{kramkov.sirbu.06}
D.~Kramkov and M.~S{\^{\i}}rbu.
\newblock Sensitivity analysis of utility-based prices and risk-tolerance
  wealth processes.
\newblock {\em Ann. Appl. Probab.}, 16(4):2140--2194, 2006.

\bibitem{kramkov.sirbu.07}
D.~Kramkov and M.~S{\^{\i}}rbu.
\newblock Asymptotic analysis of utility-based hedging strategies for small
  number of contingent claims.
\newblock {\em Stoch. Process. Appl.}, 117(11):1606--1620, 2007.

\bibitem{bouchaud.al.12}
J.~Lataillade, C.~Deremble, M.~Potters, and J.~Bouchaud.
\newblock Optimal trading with linear costs.
\newblock {\em J. Investment Strategies}, 1(3):91--115, 2012.

\bibitem{latane.59}
H.~Latan\'e.
\newblock Criteria for choice among risky ventures.
\newblock {\em J. Polit. Econ.}, 7(4):647--651, 1959.

\bibitem{liu.07}
J.~Liu.
\newblock Portfolio selection in stochastic enviroments.
\newblock {\em Rev. Finan. Stud.}, 20(1):1--39, 2007.

\bibitem{lo.wang.00}
A.~Lo and J.~Wang.
\newblock {Trading volume: definitions, data analysis, and implications of
  portfolio theory}.
\newblock {\em Rev. Financ. Stud.}, 13(2):257--300, 2000.

\bibitem{loewenstein.00}
M.~Loewenstein.
\newblock On optimal portfolio trading strategies for an investor facing
  transactions costs in a continuous trading market.
\newblock {\em J. Math. Econom.}, 33(2):209--228, 2000.

\bibitem{lynch.balduzzi.00}
A.~Lynch and P.~Balduzzi.
\newblock Predictability and transaction costs: the impact on rebalancing rules
  and behavior.
\newblock {\em J. Finance}, 55(5):2285--2310, 2000.

\bibitem{lynch.tan.11}
A.~Lynch and S.~Tan.
\newblock Explaining the magnitude of liquidity premia: the roles of return
  predictability, wealth shocks, and state dependent transaction costs.
\newblock {\em J. Finance}, 66(4):1329--1368, 2011.

\bibitem{magill.constantinides.76}
M.~Magill and G.~Constantinides.
\newblock Portfolio selection with transaction costs.
\newblock {\em J.\ Econom.\ Theory}, 13:245--263, 1976.

\bibitem{markowitz.52}
H.~Markowitz.
\newblock {Portfolio Selection, 1952}.
\newblock {\em J. Finance}, 7(1):77--91, 1952.

\bibitem{markowitz.59}
H.~Markowitz.
\newblock {\em Portfolio Selection}.
\newblock Wiley, New York, 1959.

\bibitem{martin.12}
R.~Martin.
\newblock Optimal trading under proportional transaction costs.
\newblock {\em RISK}, August, 2014.

\bibitem{martin.schoeneborn.11}
R.~Martin and T.~Sch\"oneborn.
\newblock Mean reversion pays, but costs.
\newblock {\em RISK}, February:96--101, 2011.

\bibitem{merton.69}
R.~C. Merton.
\newblock {Lifetime portfolio selection under uncertainty: the continuous-time
  case}.
\newblock {\em Rev. Econ. Statist.}, 51(3):247--257, 1969.

\bibitem{merton.71}
R.~C. Merton.
\newblock Optimum consumption and portfolio rules in a continuous-time model.
\newblock {\em J. Econom. Theory}, 3(4):373--413, 1971.

\bibitem{merton.72}
R.~C. Merton.
\newblock An analytic derivation of the efficient portfolio frontier.
\newblock {\em J. Financ. Quant. Anal.}, 7(4):1851--1872, 1972.

\bibitem{possamai.al.12}
D.~Possama\"i, H.~M. Soner, and N.~Touzi.
\newblock Homogenization and asymptotics for small transaction costs: the
  multidimensional case.
\newblock Preprint, 2012.

\bibitem{richardson.89}
H.~R. Richardson.
\newblock A minimum variance result in continuous trading portfolio
  optimization.
\newblock {\em Management Sci.}, 35(9):1045--1055, 1989.

\bibitem{rogers.04}
L.~C.~G. Rogers.
\newblock Why is the effect of proportional transaction costs
  {$O(\delta^{2/3})$}?
\newblock In {\em Mathematics of Finance}, pages 303--308. Amer. Math. Soc.,
  Providence, RI, 2004.

\bibitem{schweizer.94}
M.~Schweizer.
\newblock Approximating random variables by stochastic integrals.
\newblock {\em Ann. Probab.}, 22(3):1536--1575, 1994.

\bibitem{shreve.soner.94}
S.~E. Shreve and H.~M. Soner.
\newblock Optimal investment and consumption with transaction costs.
\newblock {\em Ann. Appl. Probab.}, 4(3):609--692, 1994.

\bibitem{soner.touzi.11}
H.~M. Soner and N.~Touzi.
\newblock Homogenization and asymptotics for small transaction costs.
\newblock {\em SIAM J. Control Optim.}, 51(4):2893--2921, 2013.

\bibitem{taksar.al.88}
M.~Taksar, M.~J. Klass, and D.~Assaf.
\newblock A diffusion model for optimal portfolio selection in the presence of
  brokerage fees.
\newblock {\em Math. Oper. Res.}, 13(2):277--294, 1988.

\bibitem{tobin.58}
J.~Tobin.
\newblock Liquidity preference as behavior towards risk.
\newblock {\em Rev. Econ. Stud.}, 25(2):65--86, 1958.

\bibitem{tobin.78}
J.~Tobin.
\newblock A proposal for international monetary reform.
\newblock {\em Eastern Econ. J.}, 4(3/4):153--159, 1978.

\bibitem{wachter.02}
J.~Wachter.
\newblock Portfolio and consumption decisions under mean-reverting returns: an
  exact solution for complete markets.
\newblock {\em J. Finan. Quant. Anal.}, 37(1):63--91, 2002.

\bibitem{whalley.wilmott.97}
A.~E. Whalley and P.~Wilmott.
\newblock An asymptotic analysis of an optimal hedging model for option pricing
  with transaction costs.
\newblock {\em Math. Finance}, 7(3):307--324, 1997.

\bibitem{zhou.li.00}
X.~Y. Zhou and D.~Li.
\newblock Continuous-time mean-variance portfolio selection: a stochastic {LQ}
  framework.
\newblock {\em Appl. Math. Optim.}, 42(1):19--33, 2000.

\end{thebibliography}

\end{document}